\documentclass{SciPost}

\hypersetup{
 colorlinks,
 linkcolor={red!50!black},
 citecolor={blue!50!black},
 urlcolor={blue!80!black}
}

\usepackage[bitstream-charter]{mathdesign}
\DeclareSymbolFont{usualmathcal}{OMS}{cmsy}{m}{n}
\DeclareSymbolFontAlphabet{\mathcal}{usualmathcal}

\fancypagestyle{SPstyle}{
\fancyhf{}
\lhead{\colorbox{scipostblue}{\bf \color{white} ~SciPost Physics }}

\fancyfoot[C]{\textbf{\thepage}}
}

\usepackage{physics}
\usepackage{booktabs}
\usepackage{amsthm}
\usepackage[capitalize]{cleveref}

\newcommand{\onlinecite}{\cite}
\newcommand{\ii}{\mathrm{i}}
\renewcommand{\Re}{\mathrm{Re}}
\renewcommand{\Im}{\mathrm{Im}}
\DeclareMathOperator{\diag}{diag}

\begin{document}

\pagestyle{SPstyle}

\begin{center}{\Large \textbf{\color{scipostdeepblue}{
Metric-induced non-Hermitian physics
\\
}}}\end{center}

\begin{center}\textbf{
Pasquale Marra
}\end{center}

\begin{center}
Department of Engineering and Applied Sciences, Sophia University, 7-1 Kioi-cho, Chiyoda-ku, Tokyo 102-8554, Japan
\\
Department of Physics \& Research and Education Center for Natural Sciences, Keio University, 4-1-1 Hiyoshi, Yokohama, Kanagawa, 223-8521, Japan
\\
Graduate School of Informatics, Nagoya University, Furo-cho, Chikusa-Ku, Nagoya, 464-8601, Japan
\\[\baselineskip]
\href{mailto:pmarra@sophia.ac.jp}{\small pmarra@sophia.ac.jp}
\end{center}

\section*{\color{scipostdeepblue}{Abstract}}
\textbf{\boldmath{%
I consider the longstanding issue of the hermiticity of the Dirac equation in curved spacetime. Instead of imposing hermiticity by adding ad hoc terms, I renormalize the field by a scaling function, which is related to the determinant of the metric, and then regularize the renormalized field on a discrete lattice. I found that, for time-independent and diagonal (or conformally flat) coordinates, the Dirac equation returns a pseudo-Hermitian (i.e., $\mathcal{PT}$-symmetric) Hamiltonian when properly regularized on the lattice. Notably, the $\mathcal{PT}$-symmetry is unbroken, ensuring a real energy spectrum and unitary time evolution. This establishes stringent conditions for the existence of complex spectra in non-Hermitian models in one dimension. Conversely, time-dependent spacetime coordinates break pseudohermiticity, yielding non-Hermitian Hamiltonians with nonunitary time evolution. Similarly, space-dependent spacetime coordinates lead to the non-Hermitian skin effect, i.e., the accumulation of localized states on the boundaries. Arguably, these non-Hermitian effects are physical: time dependence leads to local gain and loss processes and nonunitary growth or decay. Conversely, space dependence leads to the non-Hermitian skin effect with spatial decay of the fields in a preferential direction. In other words, the curvature gradients induce an imaginary gauge field on the lattice, corresponding to a drift force acting in space and time, pushing the eigenmodes to the boundaries or forcing their probability density to increase or decrease over time. Hence, temporal curvature gradients produce nonunitary gain or loss, while spatial curvature gradients correspond to the non-Hermitian skin effect, allowing for the description of these two phenomena in a unified framework. This also suggests a duality between non-Hermitian physics and spacetime deformations, framing non-Hermitian physics in purely geometric terms. This metric-induced nonhermiticity unveils an unexpected connection between the spacetime metric and non-Hermitian phases of matter.
}}

\vspace{\baselineskip}

\noindent\textcolor{white!90!black}{%
\fbox{\parbox{0.975\linewidth}{%
\textcolor{white!40!black}{\begin{tabular}{lr}%
 \begin{minipage}{0.6\textwidth}%
 {\small Copyright attribution to authors. \newline
 This work is a submission to SciPost Physics. \newline
 License information to appear upon publication. \newline
 Publication information to appear upon publication.}
 \end{minipage} & 
\end{tabular}}
}}
}


\vspace{10pt}
\noindent\rule{\textwidth}{1pt}
\tableofcontents
\noindent\rule{\textwidth}{1pt}
\vspace{10pt}

\section{Introduction}

Reconciling general relativity with quantum field theory in a mathematically and physically consistent way is an open problem in high-energy physics.
However, a well-studied semiclassical approximation of quantum gravity is the quantum field theory in curved spacetime, which is obtained when particle fields are treated as quantum-mechanical, and spacetime is treated as a classical background~\cite{mcvittie_diracs_1932,hollands_quantum_2015}.
In this framework, e.g., the Dirac equation is extended from the flat Minkowski spacetime of quantum field theory to a curved spacetime by substituting the partial derivatives with covariant ones.
Nevertheless, this theory itself is not free from apparent inconsistencies, such as the issue of the nonhermicity of the Dirac Hamiltonian in curved spacetime~\cite{parker_one-electron_1980,huang_hermiticity_2009}.
This nonhermicity is usually cured by adding extra terms that make the Hamiltonian Hermitian~\cite{huang_hermiticity_2009,nakahara_geometry_2018}, or by considering sufficiently smoothly varying metrics, which makes the non-Hermitian terms negligible.
The issue of hermiticity or nonhermiticity of the Dirac Hamiltonian is also relevant in light of the recent advances in the study of non-Hermitian quantum mechanics~\cite{yuto-ashida_non-hermitian_2020,okuma_non-hermitian_2023} and non-Hermitian quantum field theories~\cite{kawabata_topological_2021}, and on the growing field of analog gravity, i.e., the study of quantum systems that simulate curved spacetime, such as Bose-Einstein condensates~\cite{garay_sonic_2000,lahav_realization_2010,steinhauer_observation_2016,munoz-de-nova_observation_2019}, optical metamaterials~\cite{sheng_trapping_2013,bekenstein_control_2017,zhong_controlling_2018,sheng_definite_2018,drori_observation_2019}, cold atoms in optical lattices~\cite{boada_dirac_2011}, graphene~\cite{cortijo_a-cosmological_2007,juan_charge_2007,vozmediano_gauge_2008,de-juan_dislocations_2010,cortijo_geometrical_2012,iorio_quantum_2014,castro_symmetry_2018}, and Weyl semimetals~\cite{morice_synthetic_2021,konye_horizon_2022,mertens_thermalization_2022,konye_anisotropic_2023}.
These systems are described by effective Hamiltonians corresponding to a Dirac equation regularized on a lattice and are usually derived perturbatively, specifically for the condensed matter system considered~\cite{cortijo_a-cosmological_2007,juan_charge_2007,vozmediano_gauge_2008,de-juan_dislocations_2010,cortijo_geometrical_2012,iorio_quantum_2014,castro_symmetry_2018}.

Here, I consider an alternative regularization scheme and derive an effective lattice Hamiltonian by regularizing the continuum Dirac equation in curved spacetime, regardless of the specifics of the condensed matter system considered.
Generally, regularization on discrete lattices approximates the derivatives of the field with finite differences.
However, this approach leads to non-Hermitian lattice Hamiltonians, which can be made Hermitian only by imposing hermiticity by hand~\cite{boada_dirac_2011}.
Instead, here I first rewrite the Hamiltonian in terms of the derivatives of the field times a scaling function, which is related to the determinant of the metric, and then approximate these "renormalized" derivatives with finite differences.
This regularization approach returns a Hermitian or pseudo-Hermitian Hamiltonian for time-independent (static) spacetime coordinates, without arbitrarily adding extra terms to cancel out the non-Hermitian terms.
Notably, I found that these pseudo-Hermitian Hamiltonians have unbroken $\mathcal{PT}$-symmetry, corresponding to real energy spectra and unitary time evolutions.
This also establishes stringent necessary conditions for the existence of complex spectra in non-Hermitian models in one dimension.
Conversely, for spacetime coordinates that depend explicitly on the time coordinate, I found that the lattice Hamiltonian is, in general, neither Hermitian nor pseudo-Hermitian:
I argue that this nonhermicity is physical and that it corresponds to local gain and loss non-Hermitian processes on the lattice, resulting in the nonunitary time evolution of the field.
On top of that, I found that for spacetime coordinates that depend explicitly on space, the lattice Hamiltonian exhibits the so-called non-Hermitian skin effect~\cite{yuto-ashida_non-hermitian_2020,okuma_non-hermitian_2023} due to the presence of nonreciprocal couplings in the left and right directions:
Consequently, the wavefunctions decay exponentially in a preferential direction and become localized at one of the boundaries when the Hamiltonian is regularized on a finite patch of the spacetime.
The correspondence between spacetime curvature and non-Hermitian physics is encoded in the presence of an imaginary gauge field, which depends directly on the gradients of the spacetime curvature, and which is responsible for the skin effect and nonunitary time evolution.
I illustrate these general results by considering several examples, such as the Weyl~\cite{weyl_zur-gravitationstheorie_1917}, Rindler~\cite{rindler_kruskal_1966}, de~Sitter (dS), and anti-de~Sitter (AdS) metrics~\cite{de-sitter_on-the-relativity_1917,de-sitter_on-the-curvature_1918}.
Notably, I find that while the continuum theory is covariant under coordinate transformations, this covariance is lost upon lattice regularization:
Consequently, some lattice properties, including pseudohermiticity, become coordinate dependent.
Finally, I speculate on possible experimental realizations of quantum analogs of curved spacetime using condensed matter systems that simulate the lattice Hamiltonians derived here.

\section{Dirac equation in curved 1+1D spacetime} 

Let us start with the Dirac equation in curved 1+1D spacetime~\cite{mann_semiclassical_1991,morsink_black_1991,sinha_dirac_1994}
\begin{equation}\label{eq:2DDirac}
 \left[\ii\gamma^a e_a{}^\mu \partial_\mu + \frac\ii2\gamma^a \frac1{\sqrt{-g}}\partial_\mu(\sqrt{-g}\,e_a{}^\mu)-M\right]\psi=0,
\end{equation}
where $\psi$ is a two-spinor, 
$\gamma^\mu$ the flat spacetime gamma matrices 
satisfying $\{\gamma^\mu, \gamma^\nu \}=2\eta^{\mu\nu}$ with $\eta_{\mu\nu}=\diag{(1,-1)}$ the Minkowski metric,
$\sqrt{-g}$ is the 
square root 
of the determinant of the metric, and the zweibein is related to the metric by the relations $g_{\mu\nu}= e^a{}_\mu e^b{}_\nu \eta_{ab}$, and
$\eta_{ab}= e_a{}^\mu e_b{}^\nu g_{\mu\nu}$, 
with $g^{\mu\nu}=(g_{\mu\nu})^{-1}$.
In the Weyl representation, $\gamma^0=\sigma_x$ and $\gamma^1=\ii\sigma_y$.
The conserved scalar product in curved spacetime is defined as~\cite{parker_one-electron_1980,huang_hermiticity_2009} 
\begin{equation}\label{eq:scalarproduct}
(\phi,\varphi)=-\int\dd x\sqrt{-g}\phi^\dag\gamma^0e_a{}^0\gamma^a \varphi,
\end{equation}
which gives $(\phi,H\phi)-(H\phi,\phi)=\ii\int\dd x\phi^\dag\gamma^0 \partial_0(\sqrt{-g}\, e_a{}^0\gamma^a) \phi$ (see Ref.~\cite{huang_hermiticity_2009}).
This mandates that, for time-dependent metrics, the Hamiltonian is generally non-Hermitian.

\section{Conformally flat coordinates} 

Any metrics in 1+1D can be written in conformally flat coordinates as
\begin{equation}\label{eq:conformallyflatmetric}
\dd s^2=\Omega^2 \left(\dd t^2 -\dd x^2\right),
\end{equation}
with $\Omega\ge0$ the conformal factor.
Coordinate singularities correspond to points or regions in spacetime where $\Omega=0$ or $\Omega=\infty$.
The nonvanishing components of the metric tensor are $g_{00}=\Omega^2$ and $g_{11}=-\Omega^2$, square root of the determinant $\sqrt{-g}=\Omega^2$, and nonzero zweibein $e_0{}^0=\Omega^{-1}$ and $e_1{}^1=\Omega^{-1}$.
Separating time and space terms, the time evolution of the spinor field reads
\begin{equation}
 \ii \partial_0 \psi=H\psi=
\left(
- \ii \gamma_0\gamma^1 \partial_1
- \frac\ii2\gamma_0\gamma^1\frac{\partial_1\Omega}{\Omega}
- \frac\ii2\frac{\partial_0\Omega}{\Omega}
+ M\gamma_0\Omega\right)\psi
,
\end{equation}
where $\gamma_0=(\gamma^0)^{-1}$ 
($\gamma_0=\sigma_x$ and $\gamma_0\gamma^1=-\sigma_z$ in the Weyl representation),
$\partial_0\log\Omega={\partial_0\Omega}/{\Omega}$ and
$\partial_1\log\Omega={\partial_1\Omega}/{\Omega}$
are the temporal and spatial components of the spin connection,
and where the single-particle Hamiltonian density $H$ is identified as the operator acting on the spinor on the right side of the equation.
This can be conveniently recast as
\begin{equation}
 \ii \partial_0 \psi=H\psi=
- \ii \gamma_0\gamma^1 
\frac{\partial_1 \left(\sqrt{\Omega}\psi\right)}{\sqrt{\Omega}}
 - \frac\ii2\frac{\partial_0\Omega}{\Omega}\psi
 +M\Omega\gamma_0\psi
,
\label{eq:conformallyflathamiltonian}
\end{equation}
using $\partial_1 \sqrt{\Omega}=
{\partial_1 {\Omega}}/{2\sqrt{\Omega}}$.
For time-independent (static) coordinates $\partial_0\Omega=0$, this Hamiltonian is Hermitian with respect to the conserved scalar product defined in \cref{eq:scalarproduct}, which in this case gives $(\phi,H\phi)-(H\phi,\phi)=0$.

Single particle Hamiltonians can be discretized~\cite{susskind_lattice_1977} on a lattice $x=na$ by substituting the spatial derivatives with finite differences $\partial_1\psi\approx \frac1{2a}(\psi_{n+1}-\psi_{n-1})$.
However, this approach leads to non-Hermitian lattice Hamiltonians even for time-independent coordinates (see \cref{app:alternativeregularization}), unless one imposes hermiticity by hand using $\psi^\dag H\psi\to (H\psi)^\dag \psi+\psi^\dag H\psi$ (see, e.g., Eq. (15) in Ref.~\cite{boada_dirac_2011}).
Here, we can use more conveniently~\cite{marra_gauge_2025}
\begin{equation}
\partial_1\left(\sqrt{\Omega}\psi\right)
\approx 
\frac1{2a}\left(
\sqrt{\Omega_{n+1}}
\psi_{n+1}-
\sqrt{\Omega_{n-1}}
\psi_{n-1}
\right),
\label{eq:regularization}
\end{equation}
with $\Omega_n=\Omega(na,t)$ and $\psi_n=\psi(na,t)$ in \cref{eq:conformallyflathamiltonian}, yielding
\begin{equation}
H\psi_n=
- \frac{\ii}{2a} \gamma_0\gamma^1 
\left(
\sqrt{\frac{\Omega_{n+1}}{{\Omega_n}}}
\psi_{n+1}-
\sqrt{\frac{\Omega_{n-1}}{{\Omega_n}}}
\psi_{n-1}
\right)
- \frac\ii2\frac{\partial_0\Omega_n}{\Omega_n}\psi_n
+ M\Omega_n\gamma_0\psi_n
,
\label{eq:conformallyflathamiltonianlattice1}
\end{equation}
using open boundary conditions.
The lattice Hamiltonian density above recovers the continuum Hamiltonian density in the limit $a\to0$, as shown in detail in \cref{app:continuumlimit}.
Hence, the Hamiltonian is 
$\mathcal{H}=\int \dd x \psi^\dag H \psi\approx a\sum_n\psi_n^\dag H \psi_n$, which reads
\begin{align}
\mathcal{H}=&
a\sum_n
- \frac{\ii}{2a} 
\left(
\sqrt{\frac{\Omega_{n+1}}{\Omega_n}}
\psi_n^\dag
\gamma_0\gamma^1 
\psi_{n+1}-
\sqrt{\frac{\Omega_n}{\Omega_{n+1}}}
\psi_{n+1}^\dag
\gamma_0\gamma^1 
\psi_n
\right)
\nonumber\\&
 - \frac\ii2\frac{\partial_0\Omega_n}{\Omega_n}\psi_n^\dag\psi_n
+M\Omega_n\psi_n^\dag\gamma_0\psi_n,
\label{eq:conformallyflathamiltonianlattice2}
\end{align}
which is non-Hermitian on the lattice (i.e., $\mathcal{H}^\dag\neq\mathcal{H}$) for $\partial_0\Omega\neq0$ and $\partial_1\Omega\neq0$.
For flat spacetime coordinates $\partial_0\Omega=\partial_1\Omega=0$, the third term vanishes, while $\sqrt{{\Omega_{n+1}}/{\Omega_n}}=\sqrt{{\Omega_n}/{\Omega_{n+1}}}=1$ in the first and second terms, so that the Hamiltonian becomes Hermitian on the lattice (i.e., $\mathcal{H}^\dag=\mathcal{H}$) since the factor $\ii \gamma_0\gamma^1$ is anti-Hermitian.
In general, the lattice Hamiltonian can be compactly written as
\begin{equation}
\mathcal{H}=
a\sum_n
- \ii
t_n^{\rm LR}
\psi_n^\dag\gamma_0\gamma^1 \psi_{n+1}
+ \ii
t_n^{\rm RL}
\psi_{n+1}^\dag\gamma_0\gamma^1 \psi_n
 - \ii
 \delta_n
\psi_n^\dag\psi_n
+
\epsilon_n
\psi_n^\dag\gamma_0\psi_n,
\label{eq:conformallyflathamiltonianlatticecompact}
\end{equation}
or alternatively, performing the unitary tranformation $\psi_n^{}\to (-\ii)^n \psi_n^{}$,
\begin{equation}
\mathcal{H}=
a\sum_n
- 
t_n^{\rm LR}
\psi_n^\dag\gamma_0\gamma^1 \psi_{n+1}
-
t_n^{\rm RL}
\psi_{n+1}^\dag\gamma_0\gamma^1 \psi_n
 - \ii
 \delta_n
\psi_n^\dag\psi_n
+
\epsilon_n
\psi_n^\dag\gamma_0\psi_n,
\label{eq:conformallyflathamiltonianlatticecompact2ndform}
\end{equation}
with
\begin{subequations}
\label{eq:hopping}
\begin{align}
t_n^{\rm LR}&=
\frac{1}{2a}\sqrt{\frac{\Omega_{n+1}}{\Omega_n}}=
\frac{1}{2a}{e}^{\frac12\partial_1\log\Omega_{n}},
\\
t_n^{\rm RL}&=
\frac{1}{2a}\sqrt{\frac{\Omega_n}{\Omega_{n+1}}}=
\frac{1}{2a}{e}^{-\frac12\partial_1\log\Omega_{n}},
\\
\delta_n&=\frac12\frac{\partial_0\Omega_n}{\Omega_n}=\frac12\partial_0\log{\Omega_n},
\\
\epsilon_n&=M\Omega_n,
\end{align}
\end{subequations}
where $2a t_n^{\rm LR}=(2a t_n^{\rm RL})^{-1}$, and with $\partial_1f_n=f_{n+1}-f_n$ with abuse of notation.
The first two coefficients can be identified as the left-to-right and right-to-left "hopping" terms of the tight-binding model in \cref{eq:conformallyflathamiltonianlattice2} in the language of condensed matter physics.
These hopping terms are not reciprocal, i.e., the left-to-right hoppings $n\to n+1$ are not equal to the right-to-left hoppings $n\to n-1$, as long as $\partial_1\Omega\neq0$. 
For space-independent coordinates $\partial_1\Omega=0$ the hopping terms become equal and $t_n^{\rm LR}=t_n^{\rm RL}=1/2a$.
These space-dependent and nonreciprocal hopping amplitudes generalize the nonreciprocal (but uniform) hopping amplitudes of the Hatano-Nelson model~\cite{hatano_localization_1996,hatano_vortex_1997,hatano_localization_1998}.
Crucially, the nonreciprocity of the hoppings entails the nonhermiticity of the Hamiltonian.
The third coefficient instead describes non-Hermitian gain and loss processes on the lattice.
For time-independent coordinates $\partial_0\Omega=0$, the third term vanishes $\delta_n=0$.
For flat spacetime coordinates $\partial_0\Omega=\partial_1\Omega=0$, then $t_n^{\rm LR}=t_n^{\rm RL}=1/2a$ and $\delta_0=0$, so that the Hamiltonian becomes Hermitian on the lattice (i.e., $\mathcal{H}^\dag=\mathcal{H}$) since the factor $\ii \gamma_0\gamma^1$ is anti-Hermitian.
The fourth coefficient corresponds to a time-dependent and spatially-dependent on-site potential energy on the discrete lattice, which renormalizes the mass term in \cref{eq:conformallyflathamiltonianlattice2}.
The presence of coordinate singularities leads to some of the coefficients diverging or vanishing somewhere on the lattice.
Note that the lattice Hamiltonian $\mathcal {H}$ in \cref{eq:conformallyflathamiltonianlatticecompact} represents a large class of tight-binding Hamiltonians in 1D with open boundary conditions and nearest-neighbor hopping (neglecting nonlocal next-nearest hopping amplitudes), since it includes space-dependent/time-dependent hopping amplitudes, space-dependent/time-dependent on-site energies, and space-dependent/time-dependent gain/loss terms.
This lattice Hamiltonian can be constructed by deriving the tight-binding parameters from the conformal factor using \cref{eq:hopping}.
Conversely, in the simple case $\epsilon_n=0$ (giving $M=0$), one can construct a curved spacetime by deriving the temporal and spatial components of the spin connection from the tight-binding parameters of the lattice Hamiltonian
\begin{subequations}
\label{eq:inversehopping}
\begin{align}
\partial_0\log{\Omega_n}&=2\delta_n,
\\
\partial_1\log\Omega_{n}&=\log\left(t_n^{\rm LR}/t_n^{\rm RL}\right),
\end{align}
\end{subequations}
which can be integrated to obtain the conformal factor by separation of variables \linebreak $\Omega_n(t)=F(n) G(t)$ if $t_n^{\rm LR},t_n^{\rm RL}$ are time-independent and $\delta_n=\delta$ are space-independent, or using more general methods if $\partial_0\log\left(t_n^{\rm LR}/t_n^{\rm RL}\right)=2\partial_1 \delta_n$ such to satisfy the Schwarz integrability condition $\partial_0\partial_1\log\Omega_{n}
$
$
=\partial_1\partial_0\log\Omega_{n}$.
The more general case $\epsilon_n\neq0$ can be recovered by introducing a scalar potential $V_n$ (e.g., electric field) such that $\epsilon_n=M\Omega_n+QV_n$ where $Q$ is the corresponding charge.

It is natural to interpret the quantity
\begin{equation}\label{eq:imaginaryPeierls}
\phi_n=\frac\ii2\partial_1\log\Omega_{n}=\log\left(\sqrt{t_n^{\rm LR}/t_n^{\rm RL}}\right),
\end{equation}
as an imaginary Peierls phase since $t_n^{\rm LR}\propto e^{-\ii \phi_n}$ and $t_n^{\rm RL}\propto e^{\ii \phi_n}$.
This imaginary phase $\phi_n$ coincides with an imaginary gauge potential proportional to the spatial component of the spin connection, generated by local Lorentz transformations in the curved spacetime with conformal factor $\Omega$.
This gauge is a pure gauge, being the total derivative of a scalar field $\log\Omega_{n}$, analogously to a pure gauge $A^\mu=\nabla F$ which corresponds to zero magnetic field.
Note that, while a real Peierls phase corresponds to a magnetic field ($\rm{U}(1)$ gauge), this Peierls imaginary phase corresponds to geometric curvature.

\begin{table}[t]
 \caption{
Space-inversion and time-inversion symmetries of the lattice Hamiltonian and coefficients in \cref{eq:conformallyflathamiltonianlatticecompact} corresponding to conformally flat coordinates $\dd s^2=\Omega^2 (\dd t^2 - \dd x^2)$ in 1+1D with conformal factor $\Omega$.
Here, $\mathcal{P}$ represents the space-inversion symmetry (described by the unitary operator acting as $n+1\to n-1$ on the lattice site indexes), $\mathcal{T}$ represents time-reversal symmetry (described by the antiunitary complex conjugation operator), and $\mathcal{PT}$ their composition.
$\mathcal{T}$-symmetric Hamiltonians are Hermitian, while $\mathcal{PT}$-symmetric Hamiltonians are pseudo-Hermitian.
 }
 \label{tab:conformaltable1}
 \centering
 \begin{tabular}{@{} lllllllccc @{}} 
\multicolumn{10}{l}{Conformally flat coordinates / $\mathcal{P}$, $\mathcal{T}$ symmetries} \\
 \toprule
 metric &&&& coefficients 
 &&\qquad\qquad& 
 $\mathcal{PT}$ & $\mathcal{P}$ & $\mathcal{T}$ \\
 \midrule
 $\partial_0\Omega=0$, && $\partial_1\Omega=0$, && $2a t_n^{\rm LR}=2a t_n^{\rm RL}=1$, & $\delta_n=0$, & 
 & \checkmark & \checkmark&\checkmark
\\
 $\partial_0\Omega=0$, && $\partial_1\Omega\neq0$, && 
 & $\delta_n=0$, & 
 & \checkmark
\\
 $\partial_0\Omega\neq0$, && $\partial_1\Omega=0$, && $2a t_n^{\rm LR}=2a t_n^{\rm RL}=1$, & 
 &&&\checkmark
\\
 $\partial_0\Omega\neq0$, && $\partial_1\Omega\neq0$, 
 \\
 \bottomrule
 \end{tabular}
\end{table}
\begin{table}
 \caption{
Time-translation and space-translation symmetries of the lattice Hamiltonian with mass $M$ and coefficients in \cref{eq:conformallyflathamiltonianlatticecompact} corresponding to conformally flat coordinates $\dd s^2=\Omega^2 (\dd t^2 - \dd x^2)$ in 1+1D with conformal factor $\Omega$.
Here, TT represents time-translation symmetry, and ST represents space-translation symmetry.
Note that $\partial_1f_n=f_{n+1}-f_n$ with abuse of notation.
 }
 \label{tab:conformaltable2}
 \centering
 \footnotesize\addtolength{\tabcolsep}{-.4em}
 \begin{tabular}{@{} lllllllllcc @{}} 
\multicolumn{11}{l}{Conformally flat coordinates / TT, ST symmetries}\\
 \toprule
 mass && metric &&&& coefficients &&& TT & ST\\
 \midrule
 $M\neq0$, && $\partial_0\Omega=0$, && $\partial_1\Omega=0$, && $2a t_n^{\rm LR}=2a t_n^{\rm RL}=1$, & $\delta_n=0$, & 
 & \checkmark & \checkmark
\\
 $M\neq0$, && $\partial_0\Omega=0$, && $\partial_1\Omega\neq0$, && 
 & $\delta_n=0$, & 
 & \checkmark
\\
 $M\neq0$, && $\partial_0\Omega\neq0$, && $\partial_1\Omega=0$, && $2a t_n^{\rm LR}=2a t_n^{\rm RL}=1$, & 
 &&& \checkmark 
\\
 $M\neq0$, && $\partial_0\Omega\neq0$, && $\partial_1\Omega\neq0$, && 
 & 
 & 
 \\
 $M=0$, 
&& $\partial_0^2\log\Omega=\partial_0\partial_1\log\Omega=0$,
 && $\partial_1\partial_0\log\Omega=\partial_1^2\log\Omega=0$, && 
 & 
 $\epsilon_n=0$,&
 & \checkmark&\checkmark \\
 $M=0$, 
 && $\partial_0^2\log\Omega=\partial_0\partial_1\log\Omega=0$, &&
otherwise
 && 
 & $\epsilon_n=0$,&
 &\checkmark \\
 $M=0$, && 
 otherwise
 && $\partial_1\partial_0\log\Omega=\partial_1^2\log\Omega=0$, && 
 & 
 $\epsilon_n=0$,&
 && \checkmark \\
 $M=0$, && 
 otherwise
 && 
 otherwise
 &&& $\epsilon_n=0$,
 \\
 \bottomrule
 \end{tabular}
\end{table}

\begin{table}[t]
 \caption{
Examples of conformally flat coordinates $\dd s^2=\Omega^2 (\dd t^2 - \dd x^2)$ and coefficients of their corresponding lattice Hamiltonian in \cref{eq:conformallyflathamiltonianlatticecompact}, including the Weyl~\cite{weyl_zur-gravitationstheorie_1917}, Rindler~\cite{rindler_kruskal_1966}, de~Sitter (dS), and anti-de~Sitter (AdS) metrics~\cite{de-sitter_on-the-relativity_1917,de-sitter_on-the-curvature_1918}.
In the case of the time-dependent Weyl metric, one obtains a lattice Hamiltonian that is not pseudo-Hermitian, and that is moreover time-dependent and space-dependent for $M\neq0$ but time-independent with space-independent (having time-translational and space-translational symmetries) for $M=0$.
The lattice Hamiltonian in the Rindler metric is instead pseudo-Hermitian, and exhibits time-translational but no space-translational symmetry, with the exception of the massless case $M=0$.
The lattice Hamiltonian in the anti-de~Sitter metric (defined on the patch $x>0$) is pseudo-Hermitian, and exhibits time-translational but no space-translational symmetry.
The lattice Hamiltonian in the de~Sitter metric (defined on the patch $t>0$) is not pseudo-Hermitian, and exhibits space-translational and space-inversion symmetry but no time-translational symmetry.
The metric on the fourth row is a generalization of both the de~Sitter and anti-de~Sitter metrics, and it is defined on the patch $x,t>0$.
 }
 \label{tab:conformalexamplestable}
 \scriptsize\addtolength{\tabcolsep}{-.4em}
 \centering
 \begin{tabular}{@{} lllllccccc @{}} 
\multicolumn{3}{l}{Conformally flat coordinates / Examples}\\
 \toprule
& metric & coefficients &&& TT & ST & $\mathcal{PT}$ & $\mathcal{P}$ & $\mathcal{T}$ \\
 \midrule
 Weyl & $\Omega={e}^{rt+qx}$ &
 $2a t_n^{\rm LR}=(2a t_n^{\rm RL})^{-1}={e}^{\frac12qa}$, & $\delta_n=\frac{r}2$, &$\epsilon_n=M {e}^{rt+qna}$
 & if $M=0$ & if $M=0$ 
 \\
 Rindler & $\Omega={e}^{qx}$ &
 $2a t_n^{\rm LR}=(2a t_n^{\rm RL})^{-1}={e}^{\frac12qa}$, & $\delta_n=0$, &$\epsilon_n=M {e}^{qna}$
 & \checkmark & if $M=0$ &\checkmark
 \\
 & $\Omega={e}^{rt}$ &
 $2a t_n^{\rm LR}=2a t_n^{\rm RL}=1$, &$\delta_n=\frac{r}2$, & $\epsilon_n=M {e}^{rt}$
 & if $M=0$ & \checkmark && \checkmark 
 \\
 & $\Omega=(rt+qx)^{-1}$ &
 $2a t_n^{\rm LR}=(2a t_n^{\rm RL})^{-1}=\sqrt{\tfrac{rt+qna}{rt+qna+qa}}$, & $\delta_n=\frac{r}{2(rt+qna)}$, & $\epsilon_n=\frac{M}{rt+qna}$
 \\
 AdS & $\Omega=(qx)^{-1}$
 & 
 $2a t_n^{\rm LR}=(2a t_n^{\rm RL})^{-1}=\sqrt{\tfrac{n}{n+1}}$, & $\delta_n=0$, & $\epsilon_n=\frac{M}{qna}$
 & \checkmark && \checkmark\\
 dS & $\Omega=(rt)^{-1}$
 &
 $2a t_n^{\rm LR}=2a t_n^{\rm RL}=1$, & $\delta_n=-\frac{1}{2t}$, & $\epsilon_n=\frac{M}{rt}$
 && \checkmark && \checkmark \\
 \bottomrule
 \end{tabular}
\end{table}

I will now briefly elucidate the symmetry properties of the lattice Hamiltonian in \cref{eq:conformallyflathamiltonianlattice2}, in particular symmetry with respect to time translations, space translations, time reversal, and space inversion.
The lattice Hamiltonian in \cref{eq:conformallyflathamiltonianlattice2} is time-independent (invariant up to time translations) for time-independent coordinates $\partial_0\Omega=0$, and for time-dependent coordinates $\partial_0\Omega\neq0$ in the massless case $M=0$ as long as $\partial_0\log\Omega_n$ and $\partial_1\log\Omega_n$ are time-independent, i.e., as long as $\partial_0^2\log\Omega=0$ and $\partial_0\partial_1\log\Omega=0$.
Similarly, the lattice Hamiltonian is translational invariant (invariant up to space-translations) for space-independent coordinates $\partial_1\Omega=0$, and for space-dependent coordinates $\partial_1\Omega\neq0$ in the massless case $M=0$ as long as $\partial_0\log\Omega_n$ and $\partial_1\log\Omega_n$ are space-independent, i.e., as long as $\partial_1\partial_0\log\Omega=0$ and $\partial_1^2\log\Omega=0$.

Space-inversion symmetry is described by the space-inversion symmetry operator $\mathcal{P}$, defined as the unitary operator acting as $n+1\to n-1$ on the lattice site indexes.
Space-inversion symmetry requires that $t_n^{\rm RL}=t_n^{\rm LR}$, i.e., it requires space-independent coordinates.
Note that in the space-independent case, the lattice Hamiltonian is not only translational invariant but also invariant with respect to space inversion $\mathcal{P}$.
However, translational invariance alone does not mandate space inversion symmetry, such as in the massless case $M=0$ where ${\Omega_{n+1}}/{\Omega_n}$ and $(\partial_0 \Omega_n)/\Omega_n$ are space-independent and ${\Omega_{n+1}}/{\Omega_n}>1$.

Time-reversal symmetry is described by the time-reversal symmetry operator $\mathcal{T}$, defined as the antiunitary complex conjugation operator.
Time-reversal symmetry thus coincides with the property of being Hermitian.
Time-reversal symmetry requires that $t_n^{\rm RL}=t_n^{\rm LR}$ and $\delta_n=0$, i.e., it requires flat coordinates.
Although not Hermitian in general, for time-independent coordinates $\partial_0\Omega=0$ (which gives $\delta_n=0$), the lattice Hamiltonian in \cref{eq:conformallyflathamiltonianlattice2} is pseudo-Hermitian and $\mathcal{PT}$-symmetric~\cite{bender_pt-symmetric_1999,mostafazadeh_pseudo-hermiticity_2002,bender_pt-symmetric_2015}, i.e., $\mathcal{H}=\mathcal{PT}\mathcal{H}(\mathcal{PT})^{-1}=\mathcal{P}\mathcal{H}^*\mathcal{P}$, with $\mathcal{P}$ the space-inversion and $\mathcal{T}$ the time-reversal symmetry operators.
In this case, using the similarity transformation 
\begin{equation}
\label{eq:imaginarygaugetransform}
\psi_n\to
\frac1{\sqrt{\Omega_n}} \psi_n = {e}^{\ii\theta_n}\psi_n,
\qquad
\psi_n^\dag\to
\sqrt{\Omega_n} \psi_n^\dag = {e}^{-\ii\theta_n}\psi_n^\dag,
\end{equation}
which can be seen as an imaginary gauge transformation corresponding to a gauge rotation with imaginary angles~\cite{hatano_localization_1996,hatano_vortex_1997,hatano_localization_1998} $\theta_n=\frac\ii2 \log\Omega_n$, returns the Hermitian Hamiltonian
\begin{equation}
{\mathcal{H}}^\prime=
a\sum_n
- \frac{\ii}{2a} 
\left(
\psi_n^\dag
\gamma_0\gamma^1 
\psi_{n+1}-
\psi_{n+1}^\dag
\gamma_0\gamma^1 
\psi_n
\right)
+M\Omega_n\psi_n^\dag\gamma_0\psi_n,
\label{eq:conformallyflathamiltonianlatticehermitian}
\end{equation}
which is not unitarily equivalent, 
but only equivalent under the similarity transformation to \cref{eq:conformallyflathamiltonianlattice2}.
Since similarity transformation preserves the spectra, the Hamiltonian $\mathcal {H}$ is isospectral to the Hermitian Hamiltonian $\mathcal {H}'$, and therefore it has a real spectrum.
Intuitively, the imaginary gauge potential corresponds to the imaginary Peierls phase in \cref{eq:imaginaryPeierls}.
This potential is a pure gauge, being the derivative of a scalar field $\log\Omega$, which can be "gauged away" by the imaginary gauge transformation in \cref{eq:imaginarygaugetransform} under open boundary conditions:
This ensures that the spectrum remains real and the $\mathcal{PT}$-symmetry unbroken despite non-Hermitian terms.
In other words, the imaginary gauge transformation in \cref{eq:imaginarygaugetransform} corresponds to a physical change of basis to the locally flat frame where the metric appears Minkowski with $\Omega=1$, and the resulting Hamiltonian (the energy operator in the flat frame) is Hermitian.
Note that this transformation makes the hopping term reciprocal $t_n'=\sqrt{t_n^{\rm LR}t_n^{\rm RL}}=1/2a$ and leaves the on-site energy term unchanged.
The Hamiltonian ${\mathcal{H}}^\prime$ is now invariant with respect to space-inversion $\mathcal P$ and time-inversion $\mathcal T$ separately.
Moreover, it is translationally invariant for space-independent coordinates $\partial_1\Omega=0$ and for space-dependent coordinates $\partial_1\Omega\neq0$ in the massless case $M=0$.
Moreover, since the Hamiltonian $\mathcal{H}$ in \cref{eq:conformallyflathamiltonianlattice2} is isospectral to the Hermitian Hamiltonian $\mathcal{H}'$ in \cref{eq:conformallyflathamiltonianlatticehermitian} for time-independent coordinates $\partial_0\Omega=0$, this Hamiltonian $\mathcal{H}$ has real energy spectra, and thus unbroken $\mathcal{PT}$-symmetry, corresponding to unitary time evolution.
Hence, the nonreciprocity of the hoppings entails the nonhermicity of the Hamiltonian, but preserves the pseudohermicity and does not break the $\mathcal{PT}$-symmetry.
On the other hand, time-dependent coordinates break pseudohermicity and $\mathcal{PT}$-symmetry.

In general, the Hamiltonian $\mathcal {H}$ as in \cref{eq:conformallyflathamiltonianlatticecompact,eq:conformallyflathamiltonianlatticecompact2ndform} for arbitrary choices of $t_n^{\rm LR}$, $t_n^{\rm RL}$, $\epsilon_n$, \linebreak $\delta_n$ (i.e., not necessarily generated by the metric) represents a spin-1/2 tight-binding Hamiltonian with nearest-neighbor hoppings and open boundary conditions.
Any such Hamiltonian has a real spectrum as long as $\delta_n=0$ and $\epsilon_n\in\mathbb{R}$ (i.e., no gain and loss terms) \linebreak and under the additional assumption that $t_n^{\rm LR}t_n^{\rm RL}\ge0$.
Indeed, under these assumptions, \linebreak one can define a similarity transformation 
$\psi_n\to(1/{s_n}) \psi_n$,
$\psi_n^\dag\to s_n \psi_n^\dag$, 
taking $s_n=1$ \linebreak and $s_{n+1}=s_n \sqrt{t_n^{\rm LR}/t_n^{\rm RL}}$ iteratively,
which transforms the Hamiltonian into a Hermitian one \linebreak
$\mathcal{H}'=
a\sum_n
(
\ii
t_n
\psi_{n+1}^\dag\gamma_0\gamma^1 \psi_n
+h.c.)
$
$
+
\epsilon_n
\psi_n^\dag\gamma_0\psi_n,
$ with reciprocal hoppings $t_n'=\sqrt{t_n^{\rm LR}t_n^{\rm RL}}$.
In the case where the hoppings are derived from the metric via \cref{eq:imaginarygaugetransform}, one has $t_n^{\rm LR}t_n^{\rm RL} = 1/(2a)^2>0$ and $t_n^{\rm LR}/t_n^{\rm RL} = \Omega_{n+1}/\Omega_n>0$, since the conformal factor $\Omega$ is strictly positive by definition:
In this case the similarity transformation reduces to \cref{eq:imaginarygaugetransform}.
Again, since similarity transformations preserve the spectra, the Hamiltonian $\mathcal{H}$ has a real spectrum in this case.
This statement can also be derived from linear algebra alone (see \cref{app:spin12proof}).
Hence, any spin-1/2 1D lattice Hamiltonian $\mathcal {H}$ in the form \cref{eq:conformallyflathamiltonianlatticecompact,eq:conformallyflathamiltonianlatticecompact2ndform} with only nearest-neighbor hopping, real on-site potentials, hopping amplitudes $t_n^{\rm LR}t_n^{\rm RL}\ge0$, and open boundary conditions, has a real spectrum.
This statement also applies to a simpler case: 
Any spinless (or spin-diagonal) 1D lattice Hamiltonian 
$
\mathcal{H}=
a\sum_n
- 
t_n^{\rm LR}
\psi_n^\dag \psi_{n+1}
-
t_n^{\rm RL}
\psi_{n+1}^\dag \psi_n
+
\epsilon_n
\psi_n^\dag\psi_n
$ with only nearest-neighbor hopping, real on-site potentials, hopping amplitudes $t_n^{\rm LR}t_n^{\rm RL}\ge0$, and open boundary conditions, has a real spectrum (see \cref{app:spinlessproof}).

\begin{figure*}[t]
 \centering
 \includegraphics[width=\textwidth]{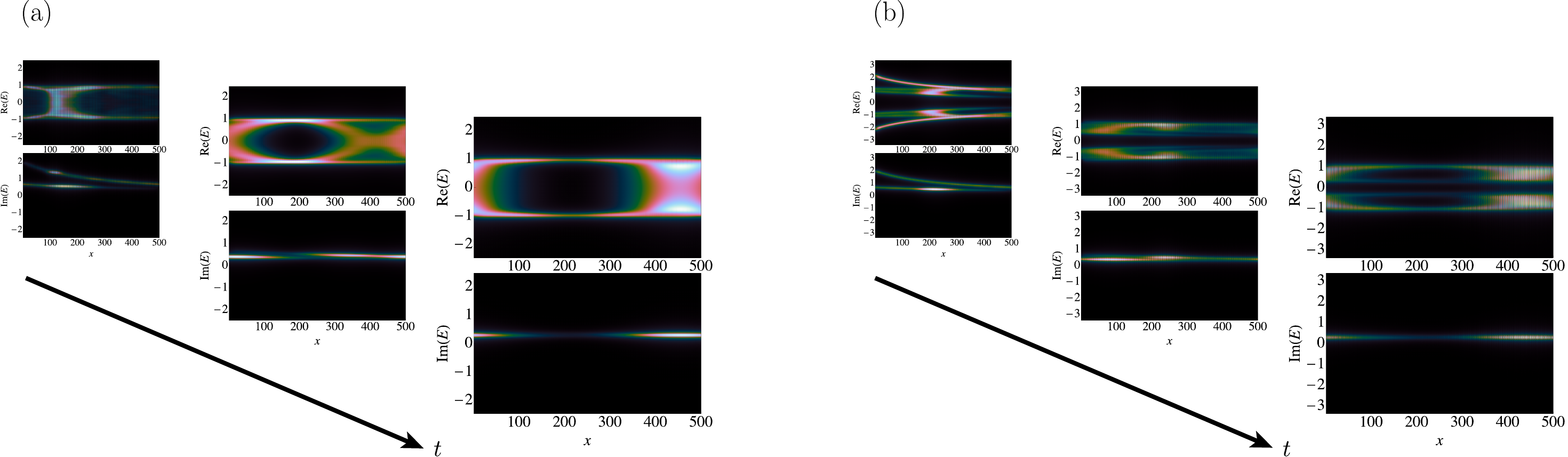}
 \caption{
Local density of states (LDOS) on the real and imaginary axes of the time-dependent and non-Hermitian lattice Hamiltonian corresponding to curved spacetime with the conformally flat metric $\dd s^2=(rt+qx)^{-1} (\dd t^2 - \dd x^2)$ [see \cref{tab:conformalexamplestable}], calculated on a finite patch as a function of the energy and position and at different time slices $t=0.25,0.5,0.75$ with $r=1$.
For time-dependent Hamiltonians, the time-dependent local density of states shown here describes the adiabatic time-evolution of the energy spectra.
Different panels correspond to the massless (a) and massive (b) cases ($M=0.5$).
This metric has a coordinate singularity at $x=t=0$.
}
 \label{fig:LDOS2}
\end{figure*}

Hence, it follows that complex energy spectra in spinless (or spin-diagonal) or spin-1/2 1D lattice Hamiltonians can arise only 
i) in the presence of nonzero gain and loss terms $\delta_n\neq0$ or $\epsilon_n\notin\mathbb{R}$, or 
ii) in the presence of next-to-nearest neighbor hopping terms, or
iii) in the presence of nearest neighbor hopping terms such that the product of hopping amplitudes in opposite directions is not a complex number or a real negative number, i.e., does not fulfill the condition $t_n^{\rm LR}t_n^{\rm RL}\ge0$, or
iv) in the presence of periodic, anti-periodic, or other generalized boundary conditions different from open boundary conditions.
Note that these conditions are necessary (but not sufficient) conditions for the existence of complex spectra.
Some counterexamples confirming the validity of these conditions are Hamiltonians with complex spectra corresponding to curved spacetime with the conformally flat metric $\dd s^2=(rt+qx)^{-1} (\dd t^2 - \dd x^2)$ [see \cref{fig:LDOS2,tab:conformalexamplestable}] or
the Hamiltonian in Ref.~\onlinecite{mardani_exceptional_2025} having imaginary (gain and loss) potentials (which do not fulfill the condition i above), 
Hamiltonians with next-to-nearest neighbor hoppings showing exceptional points as in Ref.~\onlinecite{gohsrich_exceptional_2024} (which does not fulfill the condition ii above), 
and the Hatano-Nelson model with periodic boundary conditions~\cite{hatano_localization_1996,hatano_vortex_1997,hatano_localization_1998} or with generalized boundary conditions as in Ref.~\onlinecite{rahul_controlling_2026} (which do not fulfill the condition iv above).

The properties of lattice Hamiltonians corresponding to 1+1D metrics in conformally flat coordinates are summarized in \cref{tab:conformaltable1,tab:conformaltable2}.
Some examples, including the Rindler, de~Sitter, and anti-de~Sitter metrics, are listed in \cref{tab:conformalexamplestable}.

Notice that in the Rindler metric in the massless case $M=0$, the lattice Hamiltonian coincides with that of the original Hatano-Nelson model~\cite{hatano_localization_1996,hatano_vortex_1997,hatano_localization_1998} (without disorder), which is the archetype of non-Hermitian models in condensed matter physics.
Indeed, applying the imaginary gauge transformation in \cref{eq:imaginarygaugetransform}, which becomes $\psi_n\to {e}^{-\frac12 qna}\psi_n$, $\psi_n^\dag\to {e}^{\frac12 qna}\psi_n^\dag$, returns simply $\widetilde{\mathcal{H}}=a\sum_n-\frac{\ii }{2a}\left( \psi_n^\dag \gamma_0\gamma^1 \psi_{n+1} - \psi_{n+1}^\dag \gamma_0\gamma^1 \psi_n\right)$.

\begin{figure*}[t]
 \centering
 \includegraphics[width=1\textwidth]{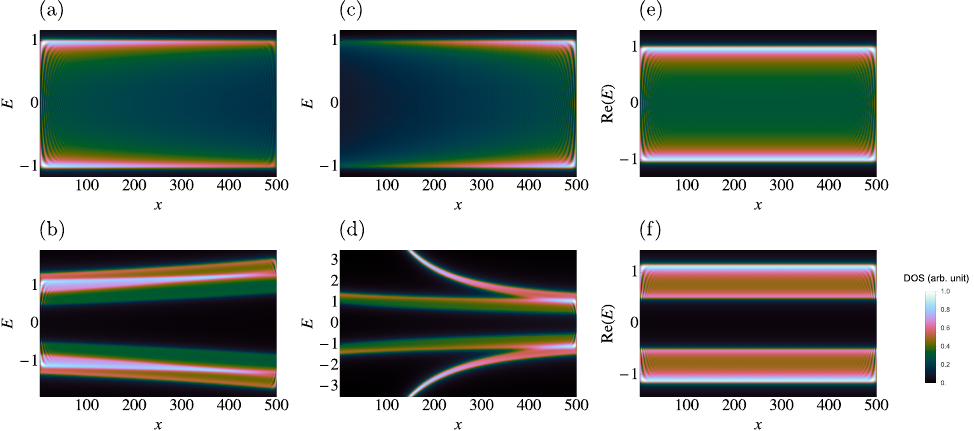}
 \caption{
Local density of states (LDOS) of the lattice Hamiltonians corresponding to the Dirac equation in curved spacetimes with conformally flat coordinates [see \cref{tab:conformalexamplestable}], calculated on a finite patch as a function of the energy and position.
Different panels correspond to:
Rindler metric in conformally flat coordinates $\dd s^2=e^{qx} (\dd t^2 - \dd x^2)$ in the massless (a) and massive (b) cases;
anti-de~Sitter metric in conformally flat coordinates $\dd s^2=(q x)^{-1} (\dd t^2 - \dd x^2)$ with coordinate singularity at $x=0$ in the massless (c) and massive (d) cases;
de~Sitter metric in conformally flat coordinates $\dd s^2=(r t)^{-1} (\dd t^2 - \dd x^2)$ with coordinate singularity at $t=0$ in the massless (e) and massive (f) cases;
In the de~Sitter case, the lattice Hamiltonian has no $\mathcal{PT}$ symmetry and thus exhibits complex energy eigenvalues, with the imaginary part of the LDOS (not shown) being time-dependent, while the real part of the LDOS remains time-independent.
The skin effect is visible for the Rindler metric [(a) and (b)] and the anti-de Sitter metric [(c) and (d)] in the massless and massive cases.
The mass in (b), (d), and (f) is $M=0.5$.
}
 \label{fig:LDOS1}
\end{figure*}

Discrete tight-binding Hamiltonians can be simulated in several different condensed matter systems, arrays of atoms~\cite{drost_topological_2017,palacio-morales_atomic-scale_2019}, quantum dots~\cite{zhao_large-area_2021}, and cold atoms in optical lattices~\cite{bloch_many-body_2008,boada_dirac_2011,minar_mimicking_2015,mula_casimir_2021}.
A common approach to probe their quantum state is to measure the electronic or atomic density of states, i.e., the density of energy levels (as a function of energy).
In some systems, one can also measure the electronic or atomic \emph{local} density of states, which is the spatially resolved density of energy levels (as a function of energy and position).
This is a direct probe of the wavefunction of the electrons or atoms in the condensed matter system.
Hence, to relate theoretical results obtained here to experimentally accessible physical properties, I show below the local density of states of the lattice Hamiltonians considered here, calculated following the method detailed in \cref{app:LDOS}.
Figure~\ref{fig:LDOS1} shows the local density of states of lattice Hamiltonians corresponding to the Dirac equation in curved spacetimes with conformally flat coordinates in the Rindler, de~Sitter, and anti-de~Sitter metrics in \cref{tab:conformalexamplestable} as a function of energy and position in the massless and massive cases.
The lattice Hamiltonian is pseudo-Hermitian for the Rindler and anti-de~Sitter metrics.
Instead, the lattice Hamiltonian is not pseudo-Hermitian for the de~Sitter metric and exhibits complex energy eigenvalues which are time-dependent (not shown).
Figure~\ref{fig:LDOS2} shows the local density of states on the real and imaginary axes of lattice Hamiltonians corresponding to the Dirac equation in curved spacetime with a conformally flat metric with coordinates $\dd s^2=(rt+qx)^{-1} (\dd t^2 - \dd x^2)$ [see \cref{tab:conformalexamplestable}] as a function of energy and position and at different time slices in the massless and massive cases.
This lattice Hamiltonian is not Hermitian and not pseudo-Hermitian, and it is time-dependent.
The density of states and the energy spectra are gapless in the massless case, while they become gapped in the massive case.

\section{Diagonal coordinates} 
				
As stated before, all metrics in 1+1D spacetime can be recast in a conformally flat form as in \cref{eq:conformallyflatmetric}.
Obviously, the geometrical structure of spacetime and all physical observables are invariant upon a change of coordinates, as a consequence of general covariance.
However, when regularizing the spatial dimension on the lattice, this invariance is broken.
Indeed, the discrete lattice introduces a preferred spatial frame and an artificial distinction between space and time. 
As a result, some physical properties can become dependent on the specific choice of the coordinates.
To show how this can happen, in this Section, I will consider coordinates corresponding to a diagonal metric tensor.
Another reason to consider these more general forms of the metric is that some of the results obtained in this Section may be easier to generalize to higher dimensions, in particular in 2+1D and 3+1D spacetimes, where spacetimes are not necessarily conformally flat.

Hence, we consider a general metric in the form 
\begin{equation}\label{eq:diagonalmetric}
\dd s^2=\alpha^2 \dd t^2 - \beta^2\dd x^2,
\end{equation}
with $\alpha,\beta\ge0$, yielding the metric tensor $g_{00}=\alpha^2$ and $g_{11}=-\beta^2$, square root of the determinant $\sqrt{-g}=\alpha\beta$, and zweibein $e_0{}^0=\alpha^{-1}$ and $e_1{}^1=\beta^{-1}$.
Coordinate singularities correspond to points or regions in spacetime where $\alpha\beta=0$ or $\alpha\beta=\infty$.
Separating time and space terms, this metric yields
\begin{equation}
 \ii \partial_0 \psi=H\psi=
\left(
- \ii \gamma_0\gamma^1\frac\alpha\beta \partial_1 
 - \frac\ii2\frac{\partial_0\beta}{\beta}
 - \frac\ii2\gamma_0\gamma^1\frac{\partial_1\alpha}{\beta}
 +M\gamma_0\alpha\right)\psi
\label{eq:diagonalhamiltonian1st}
,
\end{equation}
which can conveniently be recast as
\begin{equation}
 \ii\partial_0 \psi=H\psi=
 - 
\frac{\ii\sqrt{\alpha} }{{\beta}}
\gamma_0 \gamma^1 \partial_1 
 \left(
 \sqrt{\alpha} \psi
 \right)
- \frac\ii2\frac{\partial_0\beta}{\beta}\psi
+ M\alpha\gamma_0 \psi
,
\label{eq:diagonalhamiltonian}
\end{equation}
which is again Hermitian for time-independent coordinates.
Regularizing on the lattice using \cref{eq:regularization} for $\sqrt\alpha\psi$ yields
\begin{equation}
H\psi_n=
- \frac{\ii}{2a} \gamma_0\gamma^1 
\frac{\sqrt{\alpha_n}}{\beta_n}
\left(
\sqrt{\alpha_{n+1}}
\psi_{n+1}-
\sqrt{\alpha_{n-1}}
\psi_{n-1}
\right)
- \frac\ii2\frac{\partial_0\beta_n}{\beta_n}\psi_n
+M \alpha_n\gamma_0\psi_n
,
\label{eq:diagonalhamiltonianlattice1}
\end{equation}
with $\alpha_n=\alpha(na,t)$, $\beta_n=\beta(na,t)$, and $\psi_n=\psi(na,t)$.
Hence, the Hamiltonian is
\begin{align}
\label{eq:diagonalhamiltonianlattice2}
\mathcal{H}=
&
a\sum_n
- \frac{\ii}{2a}
\left(
\frac{\sqrt{\alpha_n\alpha_{n+1}}}{\beta_n}
\psi_n^\dag
 \gamma_0\gamma^1 
\psi_{n+1}
-
 \frac{\sqrt{\alpha_n\alpha_{n+1}}}{\beta_{n+1}}
\psi_{n+1}^\dag
 \gamma_0\gamma^1 
\psi_n
\right)
\nonumber\\&
- \frac\ii2\frac{\partial_0\beta_n}{\beta_n}\psi_n^\dag \psi_n
+ M \alpha_n\psi_n^\dag \gamma_0 \psi_n,
\end{align}
which is non-Hermitian on the lattice for $\partial_0\beta\neq0$ and $\partial_1\beta\neq0$.
Also in this case, the Hamiltonian can be written in compact form as in \cref{eq:conformallyflathamiltonianlatticecompact} with 
\begin{subequations}
\begin{align}
t_n^{\rm LR}&
=\frac{1}{2a}\frac{\sqrt{\alpha_n\alpha_{n+1}}}{\beta_n}
=\frac{1}{2a}e^{\frac12\log(\alpha_n\alpha_{n+1})-\log\beta_n}
,
\\
t_n^{\rm RL}&
=\frac{1}{2a}\frac{\sqrt{\alpha_n\alpha_{n+1}}}{\beta_{n+1}}
=\frac{1}{2a}e^{\frac12\log(\alpha_n\alpha_{n+1})-\log\beta_{n+1}}
,
\\
\delta_n&=\frac12\frac{\partial_0\beta_n}{\beta_n}=\frac12\partial_0\log{\beta_n},
\\
\epsilon_n&=M\alpha_n,
\end{align}
\end{subequations}
where the first two coefficients are the left-to-right and right-to-left hoppings, the third coefficient again describes non-Hermitian gain and loss processes, and the fourth term renormalizes the mass term as a time-dependent and spatially-dependent on-site potential energy.
The hopping terms are not reciprocal as long as $\partial_1\beta\neq0$, and become equal \linebreak $t_n^{\rm LR}=t_n^{\rm RL}=(1/2a)\sqrt{\alpha_n\alpha_{n+1}}/\beta_n$ for $\partial_1\beta=0$.
The third term vanishes $\delta_n=0$ if $\partial_0\beta=0$.
Again, the presence of coordinate singularities leads to some of the coefficients diverging or vanishing somewhere on the lattice.
These coefficients reduce to \cref{eq:hopping} for $\alpha=\beta=\Omega$.

\begin{table}[t]
 \caption{
Space-inversion and time-inversion symmetries of the lattice Hamiltonian and coefficients in \cref{eq:conformallyflathamiltonianlatticecompact} corresponding to diagonal coordinates $\dd s^2=\alpha^2 \dd t^2 - \beta^2\dd x^2$ in 1+1D.
 }
 \label{tab:diagonaltable2}
 \centering
 \begin{tabular}{@{} lllllllccc @{}} 
\multicolumn{10}{l}{Diagonal coordinates / $\mathcal{P}$, $\mathcal{T}$ symmetries} \\
 \toprule
 metric &&&& coefficients &&& $\mathcal{PT}$& $\mathcal{P}$ & $\mathcal{T}$\\
 \midrule
 $\partial_0\beta=0$, && $\partial_1\beta=0$, && $2a t_n^{\rm LR}=2a t_n^{\rm RL}=\frac{\sqrt{\alpha_n\alpha_{n+1}}}{\beta_1}$, & $\delta_n=0$, & 
 & \checkmark & \checkmark&\checkmark
\\
 $\partial_0\beta=0$, && $\partial_1\beta\neq0$, &&& $\delta_n=0$, & 
 & \checkmark
\\
 $\partial_0\beta\neq0$, && $\partial_1\beta=0$, && $2a t_n^{\rm LR}=2a t_n^{\rm RL}=\frac{\sqrt{\alpha_n\alpha_{n+1}}}{\beta_1}$, 
 &&&& \checkmark
\\
 $\partial_0\beta\neq0$, && $\partial_1\beta\neq0$, \\
 \bottomrule
 \end{tabular}
 \end{table}
 \begin{table}
 \caption{
Time-translation and space-translation symmetries of the lattice Hamiltonian with mass $M$ and coefficients in \cref{eq:conformallyflathamiltonianlatticecompact} corresponding to diagonal coordinates $\dd s^2=\alpha^2 \dd t^2 - \beta^2\dd x^2$ in 1+1D.
 }
 \label{tab:diagonaltable1}
~\\
\footnotesize\addtolength{\tabcolsep}{-.4em}
 \centering
 \begin{tabular}{@{} llllllllcc @{}} 
\multicolumn{10}{l}{Diagonal coordinates / TT, ST symmetries}\\
 \toprule
 mass && metric &&& coefficients &&& TT & ST\\
 \midrule
 $M\neq0$, && $\partial_0\alpha=\partial_0\beta=0$, && $\partial_1\alpha=\partial_1\beta=0$, & $2a t_n^{\rm LR}\!=\!\dfrac1{2a t_n^{\rm RL}}\!=\!\frac{\alpha_1}{\beta_1}$, & $\delta_n\!=\!0$, & 
 & \checkmark & \checkmark
\\
 $M\neq0$, && $\partial_0\alpha=\partial_0\beta=0$, && $\partial_1\alpha\neq0$ or $\partial_1\beta\neq0$, & 
 & $\delta_n\!=\!0$, &
 & \checkmark
\\
 $M\neq0$, && $\partial_0\alpha\neq0$ or $\partial_0\beta\neq0$, && $\partial_1\alpha=\partial_1\beta=0$, & $2a t_n^{\rm LR}\!=\!\dfrac1{2a t_n^{\rm RL}}\!=\!\frac{\alpha_1}{\beta_1}$, && 
 && \checkmark 
\\
 $M\neq0$, && $\partial_0\alpha\neq0$ or $\partial_0\beta\neq0$, && $\partial_1\alpha\neq0$ or $\partial_1\beta\neq0$, \\
 \\
 $M=0$, 
 && $\begin{aligned} \partial_0^2\log\beta=\partial_0\partial_1\log\beta&=\\\partial_0\partial_1\log(\alpha_n\alpha_{n+1})&=0,\end{aligned}$
 && $\begin{aligned} \partial_1\partial_0\log\beta=\partial_1^2\log\beta&=\\\partial_1^2\log(\alpha_n\alpha_{n+1})=0,\end{aligned}$
 && $\epsilon_n\!=\!0$,&
 & \checkmark&\checkmark \\
 $M=0$, 
 && $\begin{aligned} \partial_0^2\log\beta=\partial_0\partial_1\log\beta&=\\\partial_0\partial_1\log(\alpha_n\alpha_{n+1})&=0,\end{aligned}$
&& 
otherwise
 && $\epsilon_n\!=\!0$,&
 & \checkmark \\
 $M=0$, && 
 otherwise
 && $\begin{aligned} \partial_1\partial_0\log\beta=\partial_1^2\log\beta&=\\\partial_1^2\log(\alpha_n\alpha_{n+1})=0,\end{aligned}$
 && $\epsilon_n\!=\!0$,&
 && \checkmark \\
 $M=0$, && 
 otherwise 
 && 
 otherwise 
 \phantom{$\begin{aligned} \partial_0&\\\partial_0,\end{aligned}$}
 && $\epsilon_n\!=\!0$, &
 \\
 \bottomrule
 \end{tabular}
\end{table}

\begin{table}[h]
 \caption{
Examples of diagonal metric coordinates $\dd s^2=\alpha^2 \dd t^2 - \beta^2\dd x^2$ and coefficients of their corresponding lattice Hamiltonian in \cref{eq:conformallyflathamiltonianlatticecompact}, including the Rindler~\cite{rindler_kruskal_1966}, de~Sitter, and anti-de~Sitter metrics~\cite{de-sitter_on-the-relativity_1917,de-sitter_on-the-curvature_1918}.
Notice that the lattice Hamiltonians considered here are all pseudo-Hermitian and time-independent.
 }
 \label{tab:diagonalexamplestable}
\scriptsize\addtolength{\tabcolsep}{-.4em}
 \centering
 \begin{tabular}{@{} llllllccccc @{}} 
\multicolumn{10}{l}{Diagonal coordinates / Examples}\\
 \toprule
& metric & coefficients &&&& TT & ST & $\mathcal{PT}$ & $\mathcal{P}$ & $\mathcal{T}$\\
 \midrule
 Rindler & $\alpha={qx}$, $\beta=1$ &
 $2a t_n^{\rm LR}=2a t_n^{\rm RL}=qa\sqrt{n(n+1)}$,
 & $\delta_n=0$, & $\epsilon_n=M qna$
 && \checkmark && \checkmark& \checkmark& \checkmark\\\\
 & $\alpha={rt}$, $\beta=1$&
 $2a t_n^{\rm LR}=2a t_n^{\rm RL}=rt$,
 & $\delta_n=0$, & $\epsilon_n=M rt$
 &&&\checkmark& \checkmark & \checkmark&\checkmark\\\\
 AdS & $\alpha=\frac1\beta=\sqrt{1+(qx)^2}$&
 $2a t_n^{\rm LR}=
 \sqrt[4]{[1+(qna)^2]^3 [1+(qna+qa)^2]}
 $,
 & $\delta_n=0$, & $\epsilon_n=M \sqrt{1+(qna)^2}$
 && \checkmark && \checkmark\\
 &&
 $2a t_n^{\rm RL}=
 \sqrt[4]{[1+(qna)^2][1+(qna+qa)^2]^3}
 $,
\\
dS & $\alpha=\frac1\beta=\sqrt{1-(qx)^2}$&
 $2a t_n^{\rm LR}=
 \sqrt[4]{[1-(qna)^2]^3 [1-(qna+qa)^2]}
 $,
 & $\delta_n=0$, & $\epsilon_n=M \sqrt{1-(qna)^2}$
 && \checkmark && \checkmark\\
 &&
 $2a t_n^{\rm RL}=
 \sqrt[4]{[1-(qna)^2][1-(qna+qa)^2]^3}
 $, 
 \\
 \bottomrule
 \end{tabular}
\end{table}

I will now briefly elucidate the symmetry properties of the Hamiltonian in \cref{eq:diagonalhamiltonianlattice2}, similarly to the previous Section.
The lattice Hamiltonian is time-independent (invariant up to time translations) for time-independent coordinates $\partial_0\alpha=\partial_0\beta=0$, and for time-dependent coordinates in the massless case $M=0$ as long as $\partial_0\log{\beta_n}$, $\log(\alpha_n\alpha_{n+1})-2\log\beta_n$, and $\log(\alpha_n\alpha_{n+1})-2\log\beta_{n+1}$ are time-independent, i.e., as long as $\partial_0^2\log{\beta}=0$, $\partial_0\partial_1\log\beta=0$, \linebreak and $\partial_0\partial_1\log(\alpha_n\alpha_{n+1})=0$.
Similarly, the lattice Hamiltonian is translational invariant \linebreak (invariant up to space translations) for space-independent coordinates $\partial_1\alpha=\partial_1\beta=0$, \linebreak and for space-dependent coordinates in the massless case $M=0$ as long as $\partial_0\log{\beta_n}$, \linebreak $\log(\alpha_n\alpha_{n+1})-2\log\beta_n$, and $\log(\alpha_n\alpha_{n+1})-2\log\beta_{n+1}$ are space-independent, i.e., as long as $\partial_1\partial_0\log{\beta}=0$, $\partial_1^2\log\beta=0$, and $\partial_1^2\log(\alpha_n\alpha_{n+1})=0$~\footnote{
If $\log(\alpha_n\alpha_{n+1})-2\log\beta_n$, and $\log(\alpha_n\alpha_{n+1})-2\log\beta_{n+1}$ are space-independent (or time-independent), then their difference $\log\beta_{n+1}-\log\beta_n=\partial_1\beta_n$ must be space-independent (or time-independent).
Moreover, under the same assumptions $\log(\alpha_{n+1}\alpha_{n+2})-2\log\beta_{n+1}$ is space-independent (or time-independent), then $\log(\alpha_{n+1}\alpha_{n+2})-\log(\alpha_n\alpha_{n+1})=\partial_1\log(\alpha_n\alpha_{n+1})$ must be space-independent (or time-independent).
}.
Space-inversion symmetry $\mathcal P$ requires that $t_n^{\rm RL}=t_n^{\rm LR}$, i.e., $\partial_1\beta=0$.
Conversely, time-reversal symmetry $\mathcal T$ requires that $t_n^{\rm RL}=t_n^{\rm LR}$ and $\delta_n=0$, i.e., $\partial_0\beta=\partial_1\beta=0$.

For $\partial_0\beta=\partial_1\beta=0$ the Hamiltonian becomes
\begin{equation}
\mathcal{H}=
a\sum_n
- \frac{\ii}{2a}
\frac{\sqrt{\alpha_n\alpha_{n+1}}}
{\beta_1}
\left(
\psi_n^\dag
 \gamma_0\gamma^1 
\psi_{n+1}
-
\psi_{n+1}^\dag
 \gamma_0\gamma^1 
\psi_n
\right)
+ M \alpha_n\psi_n^\dag \gamma_0 \psi_n,
\label{eq:HamiltonianRindlerlike}
\end{equation}
which is Hermitian, even when the metric coordinates are time-dependent $\partial_0\alpha\neq0$.
We stress the fact that in the case $\partial_0\beta=\partial_1\beta=0$, the regularization approach in \cref{eq:regularization} leads to a Hermitian lattice Hamiltonian in this case, while the more standard regularization (finite differences of the wavefunction) leads to non-Hermitian lattice Hamiltonians even for time-independent coordinates unless $\partial_0\beta=\partial_1\beta=0$ and $\partial_1\alpha=0$, as shown in \cref{app:alternativeregularization}.
Moreover, the Hamiltonian exhibits both time-inversion and space-inversion symmetry since 
\begin{equation}
\label{eq:hoppingdiagonal}
t_n^{\rm LR}=t_n^{\rm RL}=\frac{1}{2a}\frac{\sqrt{\alpha_n\alpha_{n+1}}}{\beta_1},
\end{equation}
which can be identified as the hopping term of a tight-binding model and is proportional to the geometric average of the square root of the determinant of the metric on two contiguous lattice sites.
Introducing
\begin{equation}\label{eq:wn2}
\alpha={e}^{-d(x)},
\qquad
\alpha_n={e}^{-d_n},
\end{equation}
it is natural to interpret the quantity $(d_n+d_{n+1})/2$ as the distance (in some characteristic units) between the lattice sites $n$ and $n+1$, at least on patches of the spacetime where $d(x)=-\log(\alpha)\ge0$.
Generally, hopping amplitudes in condensed matter describe overlap integrals between modes localized on a single lattice site (such as Wannier functions).
These overlap integrals typically scale exponentially with the distance as $t\propto{e}^{-d_n/l}$, where $d_n$ is the distance between the modes and $l$ is a characteristic length scale describing the localization of the modes (e.g., end modes in topological insulators or superconductors or atomic modes in optical lattices).
In these systems, the coefficient $\epsilon_n=M\alpha_n$ corresponds to a spatially-dependent (and possibly time-dependent) on-site energy potential.
In this case, coordinate singularities $\alpha=0$ have a straightforward physical interpretation on the lattice:
The hopping coefficient vanishes while the distance $d_n$ diverges at the singularity.
Conversely, for coordinate singularities $\alpha=\infty$, the hopping coefficient diverges at the singularity.

For time-independent coordinates and for time-dependent coordinates with $\partial_0\beta=0$ (but not necessarily space-independent), the lattice Hamiltonian in \cref{eq:diagonalhamiltonianlattice2} is pseudo-Hermitian with unbroken $\mathcal{PT}$-symmetry, real spectra, and unitary time-evolution.
Indeed in this case, using the similarity transformation
\begin{equation}
\label{eq:imaginarygaugetransform-diagonal}
\psi_n\to
\frac1{\sqrt{\beta_n}} \psi_n = {e}^{\ii\theta_n}\psi_n,
\qquad
\psi_n^\dag\to
\sqrt{\beta_n} \psi_n^\dag = {e}^{-\ii\theta_n}\psi_n^\dag,
\end{equation}
that is, a gauge transformation with imaginary angles $\theta_n=\frac\ii2 \log\beta_n$, returns the Hermitian Hamiltonian
\begin{equation}
{\mathcal{H}}^\prime=
a\sum_n
- \frac{\ii}{2a}
 \sqrt{\frac{{\alpha_n \alpha_{n+1}}}{{\beta_n\beta_{n+1}}}}
\left(
\psi_n^\dag
 \gamma_0\gamma^1 
\psi_{n+1}
-
\psi_{n+1}^\dag
 \gamma_0\gamma^1 
\psi_n
\right)
+ 
 M \alpha_n\psi_n^\dag \gamma_0 \psi_n,
\end{equation}
which is not unitarily equivalent but isospectral to \cref{eq:diagonalhamiltonianlattice2} if $\partial_0\beta_n=0$.
As in the previous case, this transformation makes the hopping term reciprocal $t_n'=({1}/{2a}) \sqrt{{{\alpha_n \alpha_{n+1}}}/{{\beta_n\beta_{n+1}}}}$ and leaves the on-site energy term unchanged.
The Hamiltonian ${\mathcal{H}}^\prime$ is invariant with respect to space-inversion $\mathcal P$ and time-inversion $\mathcal T$ separately.
Moreover, it is translationally invariant for space-independent coordinates $\partial_1\alpha=\partial_1\beta=0$ and for space-dependent coordinates in the massless case $M=0$ as long as $\partial_1(\alpha_n/\beta_n)=0$.
Note that the condition for pseudohermicity in the case of diagonal metric tensors is weaker compared to the case of conformally flat coordinates.

The properties of lattice Hamiltonians corresponding to 1+1D diagonal metric tensor are summarized in \cref{tab:diagonaltable1,tab:diagonaltable2}.
Some examples are listed in \cref{tab:diagonalexamplestable}.
Notice that the lattice Hamiltonians corresponding to the Rindler and the anti-de~Sitter spacetimes are time-independent and pseudo-Hermitian in the conformally flat coordinates in \cref{tab:conformalexamplestable} and in the diagonal coordinates chosen in \cref{tab:diagonalexamplestable}.
In particular, the lattice Hamiltonian for the Rindler metric is Hermitian in diagonal coordinates, but only pseudo-Hermitian in the conformally flat case. 
However, the lattice Hamiltonian corresponding to the de~Sitter spacetime is not time-independent and not pseudo-Hermitian in the conformally flat case, while it is time-independent and pseudo-Hermitian in the diagonal coordinates chosen in \cref{tab:diagonalexamplestable}.
This is not surprising because, as anticipated, general covariance does not necessarily hold when the Dirac equation is regularized on the lattice.
As a consequence, some physical properties, such as pseudohermicity, may depend on the choice of the coordinates.

\begin{figure*}[t]
 \centering
 \includegraphics[width=1\textwidth]{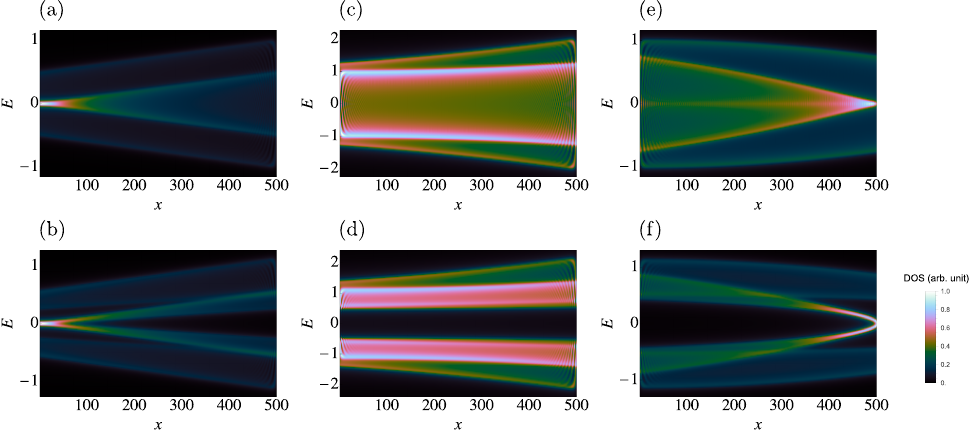}
 \caption{
Local density of states (LDOS) of the lattice Hamiltonians corresponding to the Dirac equation in curved and spacetimes with diagonal metric tensors [see \cref{tab:diagonalexamplestable}], calculated on a finite patch as a function of the energy and position.
Different panels correspond to:
Rindler metric $\dd s^2=(qx)^2 \dd t^2 - \dd x^2$ with a coordinate singularity at $x=0$ in the massless (a) and massive (b) cases;
anti-de~Sitter metric $\dd s^2=\alpha^2 \dd t^2 - \beta^{2}\dd x^2$ with $\alpha=1/\beta=\sqrt{1+(qx)^2}$ in the massless (c) and massive (d) cases;
de~Sitter metric $\dd s^2=\alpha^2 \dd t^2 - \beta^{2}\dd x^2$ with $\alpha=1/\beta=\sqrt{1-(qx)^2}$ with a coordinate singularity at $x=1/q=N$ in the massless (e) and massive (f) cases.
The mass in (b), (d), and (f) is $M=0.5$.
}
 \label{fig:LDOS3}
\end{figure*}

Figure~\ref{fig:LDOS3} shows the local density of states of lattice Hamiltonians corresponding to the Dirac equation in curved spacetimes with diagonal coordinates in the Rindler, de~Sitter, and anti-de~Sitter spacetimes in \cref{tab:diagonalexamplestable} as a function of energy and position in the massless and massive cases.

The Rindler and de~Sitter metrics exhibit coordinate singularities at $x=0$ and $x=1/q$, respectively.
These singularities correspond to localized zero-energy modes of the lattice Hamiltonian.
The presence of a singularity in the Rindler and de~Sitter metrics and the corresponding zero-energy mode mandates that the energy spectra and total density of states are always gapless (regardless of the mass term).
In particular, the local density of states is gapless everywhere in the massless case, while it becomes locally gapped in the massive case, with the exception of the singularity at $x=0$ (Rindler metric) and $x=1/q$ (de~Sitter metric).
Remarkably, the Rindler metric approximates the Schwarzschild metric near a black hole horizon.
For the anti-de~Sitter metric instead (which has no coordinate singularity), the density of states and the energy spectra are gapless in the massless case, while they become gapped in the massive case, as expected.
 
\section{Time evolution, skin effect, and nonhermiticity}

Finally, we want to unveil the physical meaning of the nonunitary time evolution ensuing from the nonhermiticity induced by the spacetime curvature. 
For conformally flat coordinates in \cref{eq:conformallyflatmetric}, the Hamiltonian in \cref{eq:conformallyflathamiltonian} can also be recast as
\begin{equation}
 \ii \gamma^0 \partial_0 \left(\sqrt{\Omega}\,\psi\right) +
 \ii \gamma^1 \partial_1 \left(\sqrt{\Omega}\,\psi\right) =
 M \Omega \left(\sqrt{\Omega}\,\psi \right)
\label{eq:niceevolution}
,
\end{equation}
where the nonunitary evolution becomes explicit.
This equation describes the unitary evolution of the field $\widetilde\psi=\sqrt{\Omega} \,\psi$ in flat spacetime, given by
$ \ii{\partial_0 \widetilde\psi}=
 - \ii
\gamma_0 \gamma^1 \partial_1 
\widetilde\psi
 +\widetilde M \gamma_0 \widetilde\psi
$,
where the renormalized mass $\widetilde M=M\Omega$ depends on the spacetime coordinates.
Regularizing on a discrete lattice $x=na$ and discrete time $t=mb$ yields
\begin{align}&
\frac{\ii}{2b}
\left(
\sqrt{\Omega_{n,m+1}}\psi_{n,m+1}
-
\sqrt{\Omega_{n,m-1}}\psi_{n,m-1}
\right)\gamma^0
\nonumber\\&
+
\frac{\ii}{2a}
 \left(
\sqrt{\Omega_{n+1,m}}
\psi_{n+1,m}-
\sqrt{\Omega_{n-1,m}}
\psi_{n-1,m}
\right)\gamma^1 
=
 M \Omega_{n,m}^{3/2}\psi_{n,m}
,
\end{align}
with $\Omega_{n,m}=\Omega(na,mb)$, and $\psi_{n,m}=\psi(na,mb)$.

In the massless case, the field in flat spacetime evolves as plane waves as $\widetilde\psi={e}^{\ii(\omega t+k x)}\chi$ with momentum $k$, energy $\omega=\pm k$, and with $\chi$ the eigenvector of $\gamma_0\gamma^1$ with eigenvalues $\pm1$.
Hence, a massless Dirac field in curved spacetime evolves like
\begin{equation}\label{eq:main}
\psi=\frac{{e}^{\ii(\omega t+k x)}}{\sqrt{\Omega}}\, \chi,
\end{equation}
which is nonunitary for $\partial_0\Omega\neq0$.
Note that the probability density is simply proportional to the inverse of the conformal factor $|\psi|^2=1/\Omega$.
In the massive case $M\neq0$, the renormalized mass depends explicitly on the spacetime coordinates, and thus the evolution of the Dirac field is not universal but depends on the details of the metric.

\begin{figure*}[t]
 \centering
	\includegraphics[width=.96\textwidth]{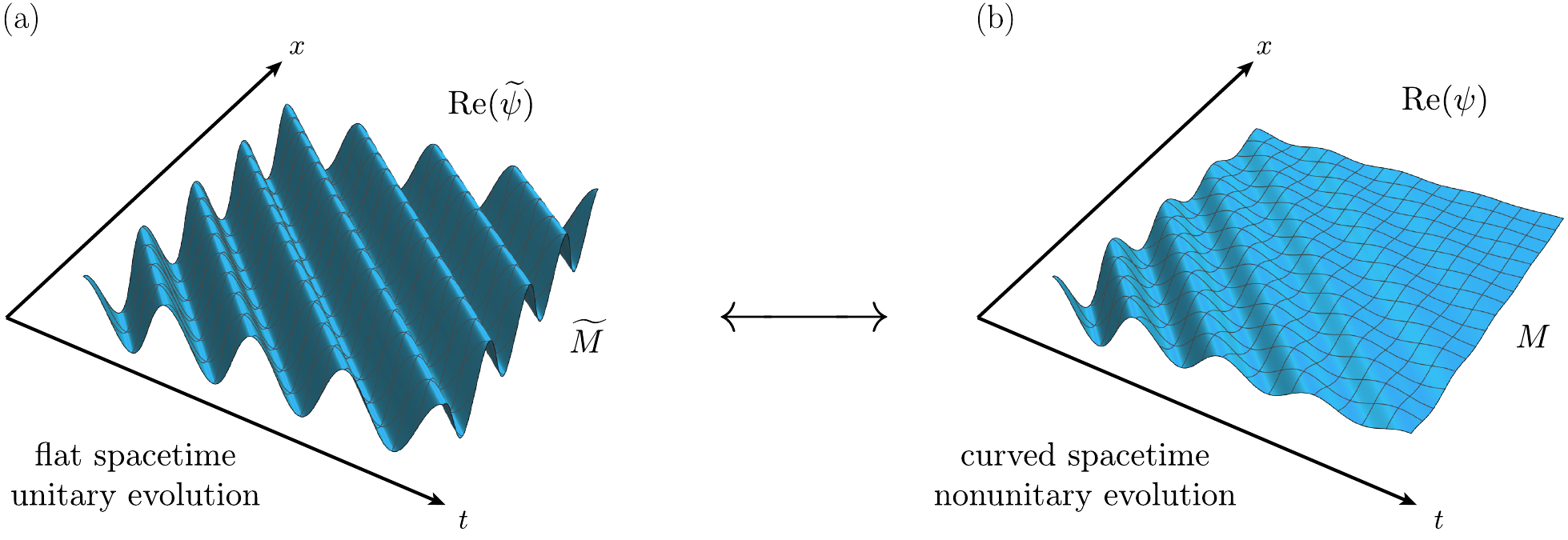}
	\caption{
The duality between the unitary evolution of the field $\widetilde\psi$ with mass $\widetilde M$ in flat spacetime and the nonunitary evolution of the field $\psi$ with mass $M$ in curved spacetime.
The relation between the two fields is determined by the metric tensor by $\widetilde\psi=\sqrt{\Omega} \,\psi$ and $\widetilde M=M\alpha$.
For simplicity, the plots show only the real part of the field in the case $\widetilde M=M=0$, with a curved spacetime corresponding to the Weyl metric with $\Omega={e}^{rt+qx}$, giving $\widetilde\psi={e}^{rt+qx} \,\psi$ with $\psi\propto {e}^{\ii (\omega t + k x)}$ (plane waves).
Note that the field $\psi$ shown above shows nonunitary time evolution for $r\neq0$ (with the probability density decaying exponentially in time) and the non-Hermitian skin effect $q\neq0$ (with the probability density decaying exponentially in space).
In the massive case $M\neq0$, the renormalized mass $\widetilde M$ depends explicitly on the spacetime coordinates.
}
 \label{fig:duality}
\end{figure*}

The $\widetilde\psi$ describes a Dirac field in flat spacetime and, as such, exhibits unitary evolution and a Hermitian Hamiltonian.
Conversely, the field $\psi$ describes a Dirac field in curved spacetime and exhibits a time evolution that is not necessarily unitary with a Hamiltonian that is not necessarily Hermitian or pseudo-Hermitian.
The relation between the two fields is determined by the metric tensor.
\Cref{fig:duality} shows the duality between the field $\widetilde\psi$ with mass $\widetilde M$ and the field $\psi$ with mass $M$, in the simplest case where the fields are massless $\widetilde M=M=0$.
Note that the duality between the Dirac field in flat spacetime and curved spacetime expressed by \cref{eq:niceevolution} is a general property of conformally flat coordinates in \cref{eq:conformallyflatmetric} that holds in the massless and massive cases.

For diagonal coordinates in \cref{eq:diagonalmetric}, the Hamiltonian in \cref{eq:diagonalhamiltonian} can also be recast as
\begin{equation}
 \ii\sqrt{\beta}\,{\partial_0 \left(\sqrt{\beta}\psi\right)}
+
{\ii }
\gamma_0 \gamma^1 \sqrt{\alpha}\, \partial_1 
 \left(
 \sqrt{\alpha} \psi
 \right)
=
 M\gamma_0 \alpha {\beta} \psi
 \label{eq:ContinuumHamiltonian2NU}
,
\end{equation}
which generalizes \cref{eq:niceevolution}.

Notably, \cref{eq:main} expresses the main result of this work, which is the relation between spacetime curvature and two main physical phenomena in non-Hermitian systems:
skin effect and nonunitary evolution~\cite{yuto-ashida_non-hermitian_2020,okuma_non-hermitian_2023}.
To understand the relation between curvature and non-Hermitian skin modes, consider a conformal metric which is monotone in space $\partial_1\Omega>0$ (or $\partial_1\Omega<0$), and assume $M=0$ (massless case) for the sake of simplicity.
In this case, the Dirac field in \cref{eq:main} gives $|\psi|^2=1/\Omega$: consequently, the probability density is also monotone in space $\partial_1|\psi|^2<0$ (or $\partial_1|\psi|^2>0$).
On a finite patch of spacetime, this mandates that the eigenmodes of the lattice Hamiltonian localize at one of the boundaries of the finite lattice: 
This is the celebrated non-Hermitian skin effect in non-Hermitian physics.
On the other hand, to understand the relation between curvature and nonunitary evolution, consider a conformal metric which is monotone in time $\partial_0\Omega>0$ (or $\partial_0\Omega<0$), with $M=0$ for the sake of simplicity.
In this case, the Dirac field $|\psi|^2=1/\Omega$ mandates that the probability density is also monotone in time $\partial_0|\psi|^2<0$ (or $\partial_0|\psi|^2>0$), i.e., it exhibits nonunitary time evolution.
Hence, the non-Hermitian skin effect corresponds to a finite space dependence of the metric.
Conversely, the nonunitary evolution corresponds to a finite time dependence of the metric.
Indeed, the two effects are dual to each other with respect to the exchange of the space and time coordinates.
Hence, the spacetime curvature allows one to describe skin effect and nonunitary time evolutions in a unified framework, as the two sides of the same coin.

Intuitively, these non-Hermitian phenomena can be described as the consequence of a persistent "drift force" acting on space and time.
Specifically, the non-Hermitian skin effect is caused by a drift force generated by the spatial component of the spin connection, which pushes the eigenmodes to the boundary.
The nonunitary time evolution is instead caused by a drift force generated by the temporal component of the spin connection, and that forces the probability density of the eigenmodes to increase or decrease over time. 
Hence, the spacetime curvature acts as a non-reciprocal pump, pushing the eigenmodes in space and time by stretching the space and time themselves.
This drift force, generated by the spin connection, is an intrinsic geometric property and can vary across the lattice and over time, for space-dependent and time-dependent curvatures, respectively.

The skin effect materializes as the accumulation of modes at one edge of the system, corresponding to a maximum of the LDOS at one edge.
This is evident for the Rindler metric for both the massless [\cref{fig:LDOS1}(a)] and massive [\cref{fig:LDOS1}(b)] cases, as well as in the anti-de Sitter metric in the massless [\cref{fig:LDOS1}(c)] and massive [\cref{fig:LDOS1}(d)] cases in conformally flat coordinates.
The nonunitary time evolution is instead visible for the time-dependent conformally flat metric in \cref{fig:LDOS2}.
The skin effect and the nonunitary time evolution in the massless case with a space-dependent and time-dependent curvature are also illustrated in \cref{fig:duality}.

\section{Discussion}

The relation between spacetime curvature and non-Hermitian effect is exemplified by considering the Weyl metric in conformally flat coordinates [see \cref{tab:conformalexamplestable}], which describes a conformal scale expansion or contraction, i.e., a transformation changing the proper distances at each point by the factor ${e}^{rt+qx}$.
In this case \cref{eq:main} gives $|\psi|^2={e}^{-(rt+qx)}$.
This corresponds to a non-Hermitian loss for time expansions $r>0$ and a non-Hermitian gain for time contractions $r<0$.
Hence, the non-Hermitian loss and gain correspond to the expansion or contraction of the spacetime background.
Note that these properties are not restricted to the Weyl metric but are valid for any conformally flat coordinates in \cref{eq:conformallyflatmetric}, where the time evolution takes the form of \cref{eq:niceevolution}, as discussed in the previous section.
The Weyl metric also allows one to exemplify the non-Hermitian skin effect in terms of spacetime curvature. 
For $q>0$, the probability density of the Dirac field increases towards the left boundary, while for $q<0$ it increases towards the right boundary of a finite spacetime patch, such that the eigenmodes become localized at one of the boundaries of the discrete lattice under open boundary conditions.
Again, these properties are not restricted to the Weyl metric but are valid for any conformally flat coordinates in \cref{eq:conformallyflatmetric}, as discussed in the previous section.

Both the Rindler and de~Sitter metrics in diagonal (but not conformally flat) coordinates defined in \cref{tab:diagonalexamplestable} exhibit a coordinate singularity, respectively at $x=0$ and $x=1/q$.
These singularities sit respectively at the lattice sites $n=0$ and $n=1/(aq)$, assuming $1/(aq)\in\mathbb{N}$ in the second case.
A defining property of these coordinate singularities is that, when a particle reaches the singularity, there is no way for the particle to escape, no matter the energy of the particle.
In the language of the lattice, this point-of-no-return property translates to the fact that the "hopping" amplitude defined in \cref{eq:hoppingdiagonal} vanishes between the lattice sites $n=0$ and $n=1$ for the Rindler metric, and between the lattice sites $n=1/(a q)-1$ and $n=1/(a q)$ for the de~Sitter metric.
Hence, the lattice points corresponding to the singularity become completely uncoupled from the rest of the lattice.
This corresponds to the presence of a localized zero-energy mode for both the massless and the massive cases.
The localized zero-energy mode is visible in \cref{fig:LDOS3}(a) and \cref{fig:LDOS3}(b) at $x=0$ for the Rindler metric, and in \cref{fig:LDOS3}(e) and \cref{fig:LDOS3}(f) at $x=N$ for the de~Sitter metric (we choose the parameter $q$ in order to match the length of the lattice $q=1/N$).
The anti-de~Sitter metric in \cref{tab:diagonalexamplestable} does not exhibit a singularity.
The energy spectrum is gapless in the massless case, and gapped in the massive case, as visible in \cref{fig:LDOS1}(c) and \cref{fig:LDOS1}(d).
The Rindler metric and generalized time-dependent Rindler-like metrics in diagonal coordinates (but not conformally flat) in \cref{tab:diagonalexamplestable} yield Hermitian Hamiltonians with unitary time evolutions.
Similarly, the de~Sitter and anti-de~Sitter metrics yield pseudo-Hermitian Hamiltonians with unbroken $\mathcal{PT}$-symmetry, again with unitary time evolutions.

There are several physical systems that simulate the Dirac equation in curved spacetime regularized on a lattice.
In essence, the regularized Hamiltonians are discrete tight-binding Hamiltonians with space-dependent and/or time-dependent hopping amplitudes and on-site energies, in combination with non-Hermitian effects such as dissipation~\cite{yuto-ashida_non-hermitian_2020,okuma_non-hermitian_2023}.
Tight-binding Hamiltonians with controllable hoppings and on-site energies are effectively simulated by arrays or lattices of atoms deposited on a surface~\cite{drost_topological_2017,palacio-morales_atomic-scale_2019}, arrays of quantum dots~\cite{zhao_large-area_2021}, cold atoms in optical lattices~\cite{bloch_many-body_2008,boada_dirac_2011,minar_mimicking_2015,mula_casimir_2021}, photonic crystals~\cite{ozawa_topological_2019}, superconducting quantum circuits~\cite{houck_on-chip_2012,anderson_engineering_2016,gu_microwave_2017}, topologically nontrivial stripes~\cite{marra_majorana_2024}, and exciton-polariton condensates in artificial lattices~\cite{valle-inclan-redondo_non-reciprocal_2024}.
Controlling the hopping amplitudes corresponds to controlling the distances and overlaps between contiguous localized states on the lattice, while on-site energies correspond to the presence of a potential.

\section{Conclusion}

In conclusion, I established here a direct correspondence between curved spacetime geometry and non-Hermitian physics by analyzing the Dirac equation in 1+1D spacetimes and its lattice regularization, unveiling the geometric origin for non-Hermitian phenomena such as nonunitary time evolutions and the skin effect.
Specifically, I have shown that for time-independent spacetime coordinates, the lattice Hamiltonians are pseudo-Hermitian and exhibit unbroken $\mathcal{PT}$-symmetry with real spectra and unitary time evolution.
In contrast, for time-dependent spacetime coordinates, the lattice Hamiltonians break $\mathcal{PT}$-symmetry, leading to nonunitary evolutions.
Similarly, I found that for space-dependent spacetime coordinates, the resulting lattice Hamiltonians exhibit asymmetric "nonreciprocal" hoppings on the lattice that localize eigenmodes at one boundary.
Hence, non-Hermitian gain or loss and nonunitary time-evolutions are induced by temporal gradients in the spacetime curvature, while spatially-asymmetric non-Hermitian terms and the skin effect are induced by spatial gradients.
Specifically, these curvature gradients induce imaginary gauge fields that act as a drift force in space and time, pushing the eigenmodes to the boundaries or forcing their probability density to increase or decrease over time.
In other words, non-Hermitian gain and loss phenomena and the skin effect can be considered the lattice counterparts of expansion and contraction of spacetime.
This provides a geometric interpretation of the non-Hermitian skin effect and of nonunitary time evolution, and unveils an unexpected connection between gravity and non-Hermitian physics.

On top of that, this work also establishes a necessary condition for the existence of complex spectra in non-Hermitian models in one dimension:
the presence of gain and loss, next-to-nearest neighbor hopping terms, nearest-neighbor hoppings such that the product of hopping amplitudes in opposite directions is a complex number or a real negative number, or boundary conditions different from open boundary conditions.

A fundamental question is whether the nonhermiticity in these models is a physical property of the Dirac field or an artifact. 
I argue that non-Hermitian phenomena, such as non-unitary time evolutions and the skin effects, are observer-dependent but physically real, in the exact same sense that inertial forces (like the centrifugal force) and Unruh radiation are observer-dependent but physically real.
Indeed, the Unruh radiation is measurable only by an observer accelerating through empty space, but not by an inertial observer at rest. 
Analogously, the non-unitary time evolution and the non-Hermitian skin effect are the measurable physical reality for an observer residing in the curved frame.
Indeed, these phenomena are observable signatures of spacetime curvature and event horizons. 
Intuitively, these effects can be understood by the fact that the conformal factor corresponds to a local scaling of the volume element by a factor $\Omega^d$. 
If the proper volume increases or decreases, one naturally expects that the local probability density is not conserved (yielding non-unitary time evolution) and/or that the probability density accumulates near event horizons (yielding the non-Hermitian skin effect).
In the context of quantum simulation of analog gravity, where the lattice represents the physical laboratory frame, non-reciprocal hoppings and gain/loss terms are thus required to faithfully reproduce the effects of gravity. 

Finally, this work suggests a broad implication on the nature of the physical reality:
Nature may not necessarily be Hermitian on a local scale, and the whole universe may, in fact, be a closed non-Hermitian system.

\newpage
\paragraph*{Acknowledgments}
I thank Naomichi Hatano and Masatoshi Sato for inspiring discussions.
This work was partially supported by the Japan Science and Technology Agency (JST) of the Ministry of Education, Culture, Sports, Science and Technology (MEXT),  JST CREST \linebreak Grant.~No.~JPMJCR19T2, the Japan Society for the Promotion of Science (JSPS) Grant-in-Aid for Early-Career Scientists Grants No.~23K13028 and No.~20K14375, Grant-in-Aid for Transformative Research Areas (A) KAKENHI Grant~No.~22H05111, and Grant-in-Aid for Transformative Research Areas (B) KAKENHI Grant~No.~24H00826.
~\\

\appendix
\crefalias{section}{appendix}

\counterwithin*{equation}{section}
\renewcommand\theequation{\thesection\arabic{equation}}

\section{Another regularization approach\label{app:alternativeregularization}}

Substituting the spatial derivatives with finite differences $\partial_1\psi\approx \frac1{2a}(\psi_{n+1}-\psi_{n-1})$ in \cref{eq:diagonalhamiltonian1st} yields
\begin{align}
 \ii\partial_0 \psi_n=
 &
 -\frac\ii{2a}\frac{\alpha_n}{\beta_n} \gamma_0\gamma^1 (\psi_{n+1}-\psi_{n-1})
 -\frac\ii2 \frac{\partial_0 \beta_n}{\beta_n} \psi_n
 -\frac\ii{4a}\left(\frac{\alpha_{n+1}}{\beta_n}-\frac{\alpha_{n-1}}{\beta_n}\right) \gamma_0\gamma^1 \psi_n
 \nonumber\\&
 +M\alpha_n\gamma_0\psi_n
,
\end{align}
and
\begin{align}
\mathcal{H}= 
&
a\sum_n
 -\frac\ii{2a} \left(
 \frac{\alpha_n}{\beta_n}
 \psi_n^\dag\gamma_0\gamma^1\psi_{n+1}
 -
 \frac{\alpha_{n+1}}{\beta_{n+1}} 
 \psi_{n+1}^\dag\gamma_0\gamma^1\psi_{n}\right)
 \nonumber\\&
 -\frac\ii2 \frac{\partial_0 \beta_n}{\beta_n} \psi_n^\dag \psi_n
 -\frac\ii{4a} \left(\frac{\alpha_{n+1}}{\beta_n} -\frac{\alpha_{n-1}}{\beta_n} \right) \psi_n^\dag\gamma_0\gamma^1 \psi_n
 +M\alpha_n\psi_n^\dag\gamma_0\psi_n
,
\end{align}
which is non-Hermitian unless $\partial_1\alpha=\partial_1\beta=0$ and $\partial_0\beta=0$, in which case one gets
\begin{equation}
\mathcal{H}= a\sum_n
 -\frac\ii{2a} 
 \frac{\alpha_1}{\beta_1}
 \left(
 \psi_n^\dag\gamma_0\gamma^1\psi_{n+1}
 -
 \psi_{n+1}^\dag\gamma_0\gamma^1\psi_{n}\right)
 +M\alpha_1\psi_n^\dag\gamma_0\psi_n
.
\end{equation}

Hence, even for simple metrics such as $\dd s^2=\alpha^2 \dd t^2 - \dd x^2$, using \cref{eq:regularization} leads to a Hermitian lattice Hamiltonian, while the more standard regularization (finite differences of the wavefunction) leads to non-Hermitian lattice Hamiltonians (with the exception of the uniform case).

\section{Continuum limit of the lattice Hamiltonian\label{app:continuumlimit}}

To rigorously establish the physical validity of the discretization scheme, I demonstrate here that the discrete Hamiltonian in \cref{eq:conformallyflathamiltonianlattice1} recovers the continuum Hamiltonian in \cref{eq:conformallyflathamiltonian} as the lattice spacing $a \to 0$. 
Let us focus on the spatial hopping term of the lattice Hamiltonian for a conformally flat metric, which reads
\begin{equation}
H_{\text{hop}}\psi_n = 
- \frac{\ii}{2a} \gamma_0\gamma^1 \left( \sqrt{\frac{\Omega_{n+1}}{\Omega_n}} \psi_{n+1} - \sqrt{\frac{\Omega_{n-1}}{\Omega_n}} \psi_{n-1} \right).
\end{equation}
By Taylor expanding at the first order, one obtains
\begin{equation}
\psi_{n\pm 1} \approx \psi \pm a \partial_1 \psi , 
\qquad
\Omega_{n\pm 1} \approx \Omega \pm a \partial_1 \Omega.
\end{equation}
where $x = na$.
The ratio of the conformal factors can be expanded as
\begin{equation}
\sqrt{\frac{\Omega_{n\pm 1}}{\Omega_n}} \approx \sqrt{\frac{\Omega \pm a \partial_1 \Omega}{\Omega}} = \sqrt{1 \pm a \frac{\partial_1 \Omega}{\Omega}} \approx 1 \pm \frac{a}{2} \frac{\partial_1 \Omega}{\Omega},
\end{equation}
using $(1+\epsilon)^{1/2} \approx 1 + \frac12\epsilon$.
Substituting these expansions into the discrete hopping term, one obtains
\begin{equation}
H_{\text{hop}}\psi_n \approx - \frac{\ii}{2a} \gamma_0\gamma^1 \left[ \left( 1 + \frac{a}{2} \frac{\partial_1 \Omega}{\Omega} \right) (\psi + a \partial_1 \psi) - \left( 1 - \frac{a}{2} \frac{\partial_1 \Omega}{\Omega} \right) (\psi - a \partial_1 \psi) \right].
\end{equation}
By neglecting higher-order terms in $a^2$, one gets
\begin{align}
H_{\text{hop}}\psi_n &\approx - \frac{\ii}{2a} \gamma_0\gamma^1 \left[ \left( \psi + a \partial_1 \psi + \frac{a}{2} \frac{\partial_1 \Omega}{\Omega} \psi \right) - \left( \psi - a \partial_1 \psi - \frac{a}{2} \frac{\partial_1 \Omega}{\Omega} \psi \right) \right] \nonumber \\
&= - \frac{\ii}{2a} \gamma_0\gamma^1 \left[ 2a \partial_1 \psi + a \frac{\partial_1 \Omega}{\Omega} \psi \right] 
= - \ii \gamma_0 \gamma^1 \left( \partial_1 + \frac{1}{2} \frac{\partial_1 \Omega}{\Omega} \right) \psi,
\end{align}
which coincides with the spatial component of the continuum Hamiltonian in \cref{eq:conformallyflathamiltonian}.
Combining this with the mass term and the time-dependent term from the original Hamiltonian, the full operator yields
\begin{equation}
\lim_{a \to 0} H = - \ii \gamma_0\gamma^1 \frac{\partial_1 \left(\sqrt{\Omega}\psi\right)}{\sqrt{\Omega}} - \frac{\ii}{2} \frac{\partial_0\Omega}{\Omega} + M \Omega \gamma_0,
\end{equation}
which is identical to the continuum Hamiltonian in \cref{eq:conformallyflathamiltonian}. This derivation proves that the nonreciprocal hopping amplitudes $t_n^{\rm LR}$ and $t_n^{\rm RL}$ are not phenomenological parameters, but physically capture the geometric effects of the spacetime curvature (vielbein and spin connection) on the lattice.

\section{Calculation of the local density of states (LDOS)\label{app:LDOS}}

To calculate the local density of states (LDOS) of a given lattice Hamiltonian with real energy spectrum as a function of energy, space, and time, one can first diagonalize the Hamiltonian in position basis and then calculate the LDOS via
\begin{equation}
\mathrm{LDOS}(x,t,E)= 
\sum_j \abs{\expval{\hat{x}}{\psi_j(t)}}^2
\delta(E_j(t)-E),
\end{equation}
where $E_j(t)$ is the $j$-th energy eigenvalue and $\ket{\psi_j(t)}$ the corresponding eigenstate at time $t$, $\hat{x}$ the position operator, and where the delta function is approximated by
\begin{equation}
\delta(x)=\frac1\pi\,\Im\left(\frac{1}{x-\ii \Gamma}\right),
\end{equation}
by taking a conveniently small $\Gamma\to0^+$.
For time-dependent Hamiltonians, the LDOS describes the adiabatic evolution of the spectra in time.
These equations apply to Hamiltonians with real energy eigenvalues, i.e., Hermitian or pseudo-Hermitian Hamiltonians with unbroken $\mathcal{PT}$-symmetry.

For non-Hermitian Hamiltonians with complex energy spectrum, one can consider the LDOS as a function of the real and imaginary parts of the energy~\cite{feinberg_spectral_1999} by taking
\begin{equation}
\mathrm{LDOS}(x,t,E) = 
\sum_j{\abs{\expval{\hat{x}}{\psi_j(t)}}^2}
\delta(\Re(E_j(t)-E))
\delta(\Im(E_j(t)-E))
,
\end{equation}
where $E_j(t)$ is the $j$-th complex energy right eigenvalue and $\ket{\psi_j(t)}$ the corresponding right eigenstate at time $t$.
On the real line, integrating on the imaginary line, one gets
\begin{equation}
\mathrm{rLDOS}(x,t,E) = 
\sum_j
{\abs{\expval{\hat{x}}{\psi_j(t)}}^2}
\delta(\Re(E_j(t)-E))
,
\end{equation}
which recovers the usual expression valid for Hermitian Hamiltonians.
Conversely, on the imaginary line, integrating on the real line yields
\begin{equation}
\mathrm{iLDOS}(x,t,E) = 
\Im\sum_j
{\abs{\expval{\hat{x}}{\psi_j(t)}}^2}
\delta(\Im(E_j(t)-E))
.
\end{equation}
These equations apply to Hamiltonians with complex energy eigenvalues, i.e., to pseudo-Hermitian Hamiltonians with no $\mathcal{PT}$-symmetry, or Hamiltonians that are neither Hermitian nor pseudo-Hermitian.

In all cases considered, the calculations are performed on a finite lattice of $500$ sites with open boundary conditions and lattice parameter $a=1$, taking $q=1/N$, $M=0$ in the massless and $M=0.5$ in the massive case, respectively.
The LDOS is plotted in arbitrary units, normalized to its maximum value (separately on the real and imaginary line), and using the "cubehelix" color scheme~\cite{green_a-colour_2011}.

\section{Necessary conditions for the existence of complex spectra in spinless 1D tight-binding Hamiltonians\label{app:spinlessproof}}

\newtheorem{theorem}{Proposition}

Here, I give a mathematical derivation of the fact that any spinless 1D lattice Hamiltonian $
\mathcal{H}=
a\sum_n
- 
t_n^{\rm LR}
\psi_n^\dag \psi_{n+1}
-
t_n^{\rm RL}
\psi_{n+1}^\dag \psi_n
+
\epsilon_n
\psi_n^\dag\psi_n
$ 
with nearest-neighbor hoppings and open boundary conditions has a real spectrum as long as $\epsilon_n\in\mathbb{R}$ (i.e., no gain and loss terms) and under the additional assumption that $t_n^{\rm LR}t_n^{\rm RL}\ge0$.
In matrix form, such a Hamiltonian can be written as
\begin{equation}
H = \begin{pmatrix} 
\epsilon_1 & t_1^{\rm LR} & 0 & \dots \\
t_1^{\rm RL} & \epsilon_2 & t_2^{\rm LR} & \dots \\
0 & t_2^{\rm RL} & \epsilon_3 & \ddots \\
\vdots & \vdots & \ddots \\
\end{pmatrix}.
\end{equation}
The reality of the spectra follows from the following proposition, which can be found in Ref.~\onlinecite{horn_matrix_2012_page174}:
\begin{theorem}
Let $A$ be a complex $N\times N$ tridiagonal matrix of the form:
\[
A = \begin{pmatrix} 
d_1 & u_1 & 0 & \dots \\
l_1 & d_2 & u_2 & \dots \\
0 & l_2 & d_3 & \ddots \\
\vdots & \vdots & \ddots & \ddots \\
\end{pmatrix},
\]
with $d_n, u_n, l_n \in\mathbb{C}$.
If $d_n\in\mathbb{R}$ and $u_n l_n \ge 0$ for all $n$, all eigenvalues of $A$ are real.
\end{theorem}

\begin{proof}
For completeness, I give here the full proof of this proposition, which is left as an exercise for the reader in Ref.~\onlinecite{horn_matrix_2012_page174}.
Let us consider the similarity transformation $S = PAP^ {-1}$, and construct $P$ in such a way that the transformed matrix $S$ is Hermitian: Since $S$ and $A$ are isospectral (because they are connected by a similarity transformation), then their spectrum must be real.
Take $P = \text{diag}(p_1, p_2, \dots, p_N)$ where 
\[
p_1 = 1, \qquad p_{n+1} = p_n \sqrt{u_n/l_n}.
\] 
If $l_n u_n > 0$, then $u_n/l_n$ is positive and $p_n$ is positive. 
The nonzero entries of $S$ are real numbers:
\begin{align*} 
s_{n, n} &= \frac{p_{n}}{p_n}d_n = d_n, \\
s_{n+1, n} &= \frac{p_{n+1}}{p_n}l_n = \sqrt{l_n u_n}, \\
s_{n, n+1} &= \frac{p_n}{p_{n+1}}u_n = \sqrt{l_n u_n}. 
\end{align*}
Since $S$ is real and symmetric, its eigenvalues are real. Because $A$ is similar to $S$, $A$ also has real eigenvalues. 
If $u_n l_n \geq 0$ instead, one can consider a matrix $A(\epsilon)$ where $u_n(\epsilon) = u_n + \epsilon$ and $l_n(\epsilon) = l_n + \epsilon$ for $\epsilon > 0$.
Now, one has $u_n(\epsilon)l_n(\epsilon)>0$ and thus all eigenvalues of $A(\epsilon)$ are real, since this reduces to the previous case.
Since the eigenvalues of a matrix are continuous functions of its entries, the eigenvalues of $A$ are the limits of the eigenvalues of $A(\epsilon)$ for $\epsilon \to 0$. 
Since the limit of a sequence of real numbers is real, all eigenvalues of $A$ are real.
\end{proof}

\section{Necessary conditions for the existence of complex spectra in spin-1/2 1D tight-binding Hamiltonians\label{app:spin12proof}}

Here, I give a purely mathematical derivation of the fact that the 1D lattice Hamiltonian $\mathcal {H}$ in the form \cref{eq:conformallyflathamiltonianlatticecompact,eq:conformallyflathamiltonianlatticecompact2ndform} for arbitrary choices of $t_n^{\rm LR}$, $t_n^{\rm RL}$, $\epsilon_n$, $\delta_n$ (spin-1/2 tight-binding Hamiltonian with nearest-neighbor hoppings and open boundary conditions) has a real spectrum as long as $\delta_n=0$ and $\epsilon_n\in\mathbb{R}$ (i.e., no gain and loss terms) and under the additional assumption that $t_n^{\rm LR}t_n^{\rm RL}\ge0$.
\Cref{eq:conformallyflathamiltonianlatticecompact2ndform} in the Weyl representation becomes
\begin{equation}
\mathcal{H}=
a\sum_n
t_n^{\rm LR}
\psi_n^\dag \sigma_z \psi_{n+1}
+
t_n^{\rm RL}
\psi_{n+1}^\dag \sigma_z \psi_n
 - \ii
 \delta_n
\psi_n^\dag\psi_n
+
\epsilon_n
\psi_n^\dag\sigma_x\psi_n, 
\end{equation}
whee the third term is zero for time-independent metrics $\delta_n=0$.
In matrix form, the Hamiltonian can be written as a block tridiagonal matrix in the form
\begin{equation}
H = \begin{pmatrix} 
D_1 & U_1 & 0 & \dots \\
L_1 & D_2 & U_2 & \dots \\
0 & L_2 & D_3 & \dots \\
\vdots & \vdots & \vdots & \ddots
\end{pmatrix}
\end{equation}
where
\begin{gather}
D_n = 
\epsilon_n\sigma_x -\ii\delta_n =
\begin{pmatrix} 
-\ii\delta_n & \epsilon_n \\
\epsilon_n & -\ii\delta_n \\
\end{pmatrix},
\nonumber\\
U_n = 
t_n^{\rm LR} \sigma_z=
\begin{pmatrix} 
t_n^{\rm LR} & 0 \\
0 & -t_n^{\rm LR} \\
\end{pmatrix},
\quad
L_n = 
t_n^{\rm RL} \sigma_z=
\begin{pmatrix} 
t_n^{\rm RL} & 0 \\
0 & -t_n^{\rm RL} \\
\end{pmatrix}.
\end{gather}
The upper and lower blocks are in the form $U_n = u_n Z$ and $L_n=l_n Z$ with $Z$ being a Hermitian matrix.
Moreover, for $\delta_n=0$ and $\epsilon_n\in\mathbb{R}$, the blocks $D_n$ are Hermitian matrices.
The reality of the spectra follows from the following proposition, which is a generalization to block matrices of the proposition in \cref{app:spinlessproof}:

\begin{theorem}
Let $A$ be a complex block tridiagonal matrix partitioned into $N \times N$ blocks, where each block is of size $M \times M$ ($M \geq 1$):
\[
A = \begin{pmatrix} 
D_1 & U_1 & 0 & \dots \\
L_1 & D_2 & U_2 & \dots \\
0 & L_2 & D_3 & \dots \\
\vdots & \vdots & \vdots & \ddots
\end{pmatrix}
,
\]
and assume that 
 i) the diagonal blocks are Hermitian: $D_n = D_n^\dag$, 
 ii) the off-diagonal blocks are $L_n = l_n Z$ and $U_n = u_n Z$, where $l_n, u_n \in \mathbb{C}$ and $Z$ is any Hermitian matrix, and 
 iii) the following condition is satisfied $l_n u_n \ge 0$ for all indexes $n$.
Under these conditions, all eigenvalues of $A$ are real.
\end{theorem}

\begin{proof}
Let us consider the similarity transformation $S = PAP^ {-1}$, and construct $P$ in such a way that the transformed matrix $S$ is Hermitian: Since $S$ and $A$ are isospectral (because they are connected by a similarity transformation), then their spectrum must be real.
We take $P = \text{diag}(p_1 I, p_2 I, \dots, p_n I)$, with $I$ is the $M \times M$ identity matrix and $p_n>0$. 
Since each block $P_n = p_n I$ is a multiple of the identity, it commutes with any complex or real matrix, and thus $P_n D_n = D_n P_n$. 
The nonzero blocks of $S = P A P^{-1}$ are:
\begin{align*}
S_{n,n} &= (p_n I) D_n (p_n I)^{-1} = D_n, \\
S_{n+1,n} &= (p_{n+1} I) (l_n Z) (p_n I)^{-1} = \left( \frac{p_{n+1}}{p_n} l_n \right) Z, \\
S_{n,n+1} &= (p_n I) (u_n Z) (p_{n+1} I)^{-1} = \left( \frac{p_n}{p_{n+1}} u_n \right) Z. 
\end{align*}
The matrix $S$ is Hermitian if $S_{n,n} = S_{n,n}^\dag$ and $S_{n+1,n} = S_{n,n+1}^\dag$.
The diagonal blocks $S_{n,n} = D_n$ are Hermitian by assumption.
The condition $S_{n+1,n} = S_{n,n+1}^\dag$ is satisfied if
\[ \frac{p_{n+1}}{p_n} l_n Z = \left( \frac{p_n}{p_{n+1}} u_n Z \right)^\dagger = \frac{p_n}{p_{n+1}} u_n Z, \]
given that $Z = Z^\dagger$ by definition.
This condition is satisfied if:
\[ \left(\frac{p_{n+1}}{p_n}\right)^2 = \frac{u_n}{l_n}. \]
If $l_n u_n > 0$, then $u_n/l_n$ is positive and $p_n$ is positive. 
Hence, one can define 
\[
p_1 = 1, \qquad p_{n+1} = p_n \sqrt{u_n/l_n},
\] 
iteratively, ensuring $P$ is a real diagonal matrix with only positive entries on the diagonal.
Under this transformation, $S$ is a Hermitian matrix, and consequently all eigenvalues are real. 
Since $A$ is similar to $S$, they share the same spectrum. 
Therefore, all eigenvalues of $A$ are real. 
If $u_n l_n \geq 0$ instead, one can consider a matrix $A(\epsilon)$ where $u_n(\epsilon) = u_n + \epsilon$ and $l_n(\epsilon) = l_n + \epsilon$ for $\epsilon > 0$.
Now, one has $u_n(\epsilon)l_n(\epsilon)>0$ and thus all eigenvalues of $A(\epsilon)$ are real, since this reduces to the previous case.
Since the eigenvalues of a matrix are continuous functions of its entries, the eigenvalues of $A$ are the limits of the eigenvalues of $A(\epsilon)$ for $\epsilon \to 0$. 
Since the limit of a sequence of real numbers is real, all eigenvalues of $A$ are real.
\end{proof}

The reality of the spectrum is not guaranteed in the following cases:
i) In the presence of gain and loss: in this case, the diagonal blocks $D_n$ are not Hermitian.
ii) In the presence of next-to-nearest neighbor hopping terms.
iii) In the presence of nearest-neighbor hoppings such that the product of hopping amplitudes in opposite directions is a complex number or a real negative number: in this case, the $p_n$ are complex-valued, giving $S_{n+1,n} \neq S_{n,n+1}^\dag$.
iv) In the presence of boundary conditions different from open boundary conditions.
In the cases i) and iii), the matrix $S$ is not Hermitian anymore, and therefore does not necessarily have a real spectrum.
In the cases ii) and iv), the Hamiltonian cannot be written as a block tridiagonal matrix, and therefore the proposition above cannot be applied.


\end{document}